\documentclass[prd,twocolumn,showpacs,nofootinbib, aps,10pt,superscriptaddress]{revtex4-1}
\usepackage{bm}
\usepackage{graphicx}
\usepackage{multirow}
\usepackage{amsmath}
\usepackage{amssymb}
\usepackage{tabulary}
\usepackage{hyperref}
\usepackage{xspace}
\usepackage{verbatim}
\usepackage{mathtools}
\usepackage{booktabs}
\usepackage[dvipsnames]{xcolor}
\usepackage[caption=false]{subfig}
\usepackage{enumitem}
\usepackage{boldline}
\usepackage{epstopdf}
\usepackage{tabularx}
\usepackage{MnSymbol}
\usepackage[utf8]{inputenc}

\newcolumntype{Y}{>{\centering\arraybackslash}X}
\hypersetup{urlcolor=Blue,
	    citecolor=OliveGreen,
	    linkcolor=Blue, 
	    colorlinks=true}
\definecolor{green}{rgb}{0,0.5,0}
\definecolor{red}{rgb}{0.5,0,0}
\definecolor{blue}{rgb}{0,0,0.5}

\newcommand{\ra}[1]{\renewcommand{\arraystretch}{#1}}

\newcommand{\vmin}{v_\mathrm{min}}
\newcommand{\qhat}{\hat{\mathbf{q}}}

\newcommand{\kms}{\textrm{ km s}^{-1}}

\newcommand{\vesc}{v_{\rm esc}}
\newcommand{\erf}{\mathrm{erf}}

\begin{document}

\title{Time-integrated directional detection of dark matter}

\author{Ciaran A. J. O'Hare}
\email{ciaran.aj.ohare@gmail.com}
\affiliation{School of Physics and Astronomy, University of Nottingham, University Park, Nottingham, NG7 2RD, UK}

\author{Bradley J. Kavanagh}
\email{bradley.kavanagh@lpthe.jussieu.fr}
\affiliation{LPTHE, CNRS, UMR 7589, 4 Place Jussieu, F-75252, Paris, France}

\author{Anne M. Green}
\email{anne.green@nottingham.ac.uk}
\affiliation{School of Physics and Astronomy, University of Nottingham, University Park, Nottingham, NG7 2RD, UK}

\begin{abstract}
The analysis of signals in directional dark matter (DM) detectors typically assumes that the directions of nuclear recoils can be measured in the Galactic rest frame. However, this is not possible with all directional detection technologies. In nuclear emulsions, for example, the recoil events must be detected and measured after the exposure time of the experiment. Unless the entire detector is mounted and rotated with the sidereal day, the recoils cannot be reoriented in the Galactic rest frame. We examine the effect of this `time integration' on the primary goals of directional detection, namely: (1) confirming that the recoils are anisotropic; (2) measuring the median recoil direction to confirm their Galactic origin; and (3) probing below the neutrino floor. We show that after time integration the DM recoil distribution retains a preferred direction and is distinct from that of Solar neutrino-induced recoils.  Many of the advantages of directional detection are therefore preserved and it is not crucial to mount and rotate the detector. Rejecting isotropic backgrounds requires a factor of 2 more signal events compared with an experiment with event time information, whereas a factor of $1.5-3$ more events are needed to measure a median direction in agreement with the expectation for DM. We also find that there is still effectively no neutrino floor in a time-integrated directional experiment. However to reach a cross section an order of magnitude below the floor, a factor of $\sim8$ larger exposure is required than with a conventional directional experiment. We also examine how the sensitivity is affected for detectors with only 2D recoil track readout, and/or no head-tail measurement. As for non-time-integrated experiments, 2D readout is not a major disadvantage, though a lack of head-tail sensitivity is. 
\end{abstract}

\maketitle

\section{Introduction}
Direct detection experiments aim to detect dark matter (DM) in the form of weakly interacting massive particles (WIMPs) via nuclear recoil events~\cite{Goodman:1984dc}. The motion of the Solar system with respect to the Galactic rest frame leads to a large asymmetry in the directions of DM-induced nuclear recoils~\cite{Spergel:1987kx}. The recoil directions are tightly concentrated around the inverse of the direction of Solar motion (towards the constellation Cygnus), with the recoil rate in the forwards hemisphere being an order of magnitude larger than the reverse. Directional detection experiments, which aim to measure the directions of the nuclear recoils as well as their energies,  are therefore a powerful way of discriminating DM-induced recoils from backgrounds and establishing the Galactic origin of a signal (see e.g.~Ref.~\cite{Mayet:2016zxu} for a review). With an ideal detector the anisotropy of the recoils could be demonstrated with only of order ten signal events~\cite{Copi:1999pw,Copi:2000tv,Morgan:2004ys} and roughly 30 events would be sufficient to measure the median recoil direction~\cite{Billard:2009mf,Green:2010zm}. Directional detection also represents one of the only ways to measure the structure of the DM velocity distribution in the local Milky Way halo~\cite{Billard:2010jh,Billard:2012qu,Lee:2012pf,O'Hare:2014oxa,Kavanagh:2016xfi,Laha:2016iom,Nagao:2017yil}.
 
The sensitivity of non-directional direct detection experiments will be limited in the near future by the recently observed coherent neutrino-nucleus scattering~\cite{Akimov:2017ade} which produces nuclear recoils with energy spectra that are very similar to DM~\cite{Drukier:1986tm,Monroe:2007xp,Strigari:2009bq,Gutlein:2010tq,Harnik:2012ni,Billard:2013qya}. Neutrino- and DM-induced recoils have very different angular spectra however, which allows directional detection experiments to discriminate between them and access cross sections below the neutrino `floor' faced by non-directional experiments~\cite{Grothaus:2014hja,O'Hare:2015mda,Franarin:2016ppr}.

While directional detection is theoretically very appealing, measuring the directions of sub-100 keV nuclear recoils is practically very challenging. Research and development to date has largely been focused on low pressure gas Time Projection Chambers  (TPCs), for reviews see Refs.~\cite{Ahlen:2009ev,Battat:2016pap}. Nuclear recoils in gas produce tracks which are $\mathcal{O}({\rm mm})$ length, and their directions can be measured in a TPC by drifting the ionisation produced along the track to a time sampled readout plane~\cite{Battat:2017tux,Couturier:2016isu,Deaconu:2017vam}. Unfortunately for gaseous detectors, extremely large target volumes will be required to probe cross sections below the stringent limits set by the leading non-directional dual-phase Xenon TPCs~\cite{Akerib:2016vxi,Tan:2016zwf,Aprile:2017iyp}. 

Recently high resolution ($\sim 10$ nm) nuclear emulsions have been developed, which facilitates the possibility of a (more easily scalable) solid directional detector, such as the proposed NEWSdm experiment~\cite{Aleksandrov:2016fyr,Agafonova:2017ajg}. A nuclear emulsion experiment would have one very significant difference from a typical directional experiment: the emulsion plates are scanned at the end of an exposure. This means that the experiment measures the time-integrated recoil rate, which is less anisotropic owing to the Earth's rotational and orbital motion. This could potentially be mitigated by constantly rotating the experiment, so that its reference frame tracks the direction of Solar motion, but this is likely to be technically challenging and financially expensive~\cite{Aleksandrov:2016fyr,Agafonova:2017ajg}. 

We investigate how integrating over time affects the angular recoil spectrum and discovery reach of directional experiments.
Similar effects of time integration on directional detection have previously been studied in another context; Ref.~\cite{SnowdenIfft:1997hd} looked at the directionality of tracks in ancient mica, laid down over $\sim$ 100 Myr. Taking into account the averaging over the orbital motion of the Solar system and plate tectonic drift, they found that a  $\sim 1\%$ asymmetry in the track orientations remains. The timescales that are relevant for lab experiments ($ \sim 1$ year) are many orders of magnitude smaller than those for ancient mica so a much larger asymmetry will remain in the cases we consider here. In addition to the effects on an idealised experiment we also study the impact of time integration when combined with 2D track readout and no head-tail sensitivity. Reference~\cite{Copi:2005ya} examined the sensitivity of a 2D detector that rotates so as to track the direction of Solar motion. They found that such a detector requires only a factor of 2 more events to establish that recoils are anisotropic than an ideal 3D detector, and an order of magnitude less than a 3D detector without head-tail sense recognition. We go significantly beyond this work and explore the prospects of measuring the median recoil direction and discriminating between DM- and neutrino-induced recoils using \textit{time-integrated} 2D and 3D detectors. We also consider a wider range of detector configurations. We highlight nuclear emulsions as the exemplar of time-integrated detectors, although the issues we discuss also apply to a number of other proposals for directional experiments using, for example, DNA~\cite{Drukier:2012hj} or crystal defects~\cite{Rajendran:2017ynw}.

The quantitative results of this paper are summarised in Table~\ref{tab:results}. In Sec.~\ref{sec:news} we overview the NEWSdm experiment, before describing in Sec.~\ref{sec:eventrate} the calculation of the directional event rate and how it is affected by integrating over time. We then show how this integration affects the sensitivity of the experiment. First, in Sec.~\ref{sec:ideal}, in order to isolate the role played by time integration, we consider an idealised detector with 3D readout and head-tail sensitivity. Comparing time-integrated to standard ``Cygnus-tracking'' experiments we calculate the number of events required to establish that the recoils are anisotropic (Sec.~\ref{sec:ideal-iso}); to confirm a preferred median direction (Sec.~\ref{sec:ideal-med}); and probe cross sections below the neutrino floor (Sec.~\ref{sec:ideal-neutrinos}). Then in Sec.~\ref{sec:nonideal} we move beyond the idealised case to examine the effects of the limitations encountered in the implementation of directional detection in general, that are particularly important in the context of nuclear emulsion-based experiments: 2D track readout and the lack of head-tail sensitivity. Finally we conclude with a summary and discussion of our results in Sec.~\ref{sec:conclusions}.

\section{NEWS\lowercase{dm} experiment}
\label{sec:news}
In this paper we study the features of the time-integrated directional rate and time-integrating detectors in general.
Nuclear emulsion-based experiments are a concrete example of a proposed time-integrating directional detector. We therefore begin by overviewing the features of the nuclear emulsions WIMP search (NEWSdm)~\cite{Aleksandrov:2016fyr}, a proposed directionally-sensitive dark matter detection experiment, which motivate some of our choices of benchmark experimental parameters.

A nuclear emulsion consists of a layer of gelatin sprinkled with semiconducting crystals of a silver halide. Ionisation produced by a recoil event leads to interactions between the silver ions and electrons, subsequently forming several-nanometer sized clusters of silver atoms. After the nuclear emulsion layers are developed the specks of silver left by a recoil track increase in size to tens of nanometers~\cite{DAmbrosio:2014arr}. Potential nuclear recoil tracks can then be identified and imaged by a scanning optical microscope. Subsequently a hard X-ray microscope can be used to confirm that the candidate events are nuclear recoils and achieve higher resolution. Nuclear emulsions are excellent targets for fast particle detection, responsible for the discovery of the pion~\cite{Lattes:1947mw} and the first detection of $\nu_\mu \rightarrow \nu_\tau$ neutrino oscillations~\cite{Agafonova:2010dc}.  They have not previously been used for the detection of dark matter. A key recent development in the viability of nuclear emulsions for directional DM detection is the fabrication of Nano Imaging Trackers (NITs). These novel emulsion films have grain diameters of order tens of nm~\cite{Natsume:2007zz,Naka:2013mwa}, an order of magnitude smaller than conventional emulsions~\cite{Acquafredda:2009zz}.

While nuclear emulsions have an advantage compared to gaseous detectors, namely in scalability, they have other limitations. Firstly they are not expected to have head-tail sense recognition i.e. recoils in the directions +$\qhat$ and -$\qhat$ are indistinguishable. Secondly, events are not time-tagged. The distribution of recoil directions is therefore averaged over many rotations of the Earth, which partially washes out the directional signal.  This effect can potentially be minimised by mounting the experiment on an equatorial telescope which tracks the constellation Cygnus, towards which the Solar system is moving. However, such a telescope would make up a substantial fraction of the total experimental cost (see e.g.~Table~6 of Ref.~\cite{Aleksandrov:2016fyr}). In addition, the feasibility of such a setup has yet to be demonstrated for a large detector with the shielding required to achieve low backgrounds.

Reference~\cite{Agafonova:2017ajg} explores the discovery potential for a directional dark matter search with a nuclear emulsion-based detector. They simulate track propagation and straggling (the deviation of the recoil track from the initial nuclear recoil direction) with a 100 nm track length threshold for a mixture of target nuclei (mass fraction in \%): H (1.6), C (10.1), N (2.7), O(7.4), S(0.3), Br(32.0), Ag (44.0), I (1.9). They assume that the detector accounts for the rotation of the Earth with an equatorial mount, with the nuclear emulsion plates oriented so that the direction of Solar motion lies in the plane of the plates. 
The long-term goal of the experiment is to record 3D tracks, however for this analysis they consider only the 2D track projection into this plane. They also assume that there is no head-tail sense discrimination. The distribution of the angles between the tracks and the initial recoil directions is a Gaussian centered at $0^{\circ}$  with a width $\sigma_\theta$. 
The width has a maximum value of $\sigma_\theta \sim 30^{\circ}$ at low recoil energies ($\sim$ 40 keV) and decreases to a plateau at around $\sigma_\theta\sim 20^{\circ}$ for high recoil energies ($\sim 300$ keV).

They find that the directionality (for a detector on an equatorial mount) allows a $\sim (10-20) \%$ improvement in sensitivity over that which could be achieved by an experiment with no direction sensitivity. They also show that the discovery limit for a zero background 10 ton-year experiment with a lower energy threshold would reach the neutrino floor for xenon (as calculated by Billard~et~al.~\cite{Billard:2013qya}) where it is highest at $m_{\chi} \sim 8 \, {\rm GeV}$. However the neutrino floor (i.e.~the cross section at which neutrino backgrounds limit the sensitivity of non-directional experiments) is target dependent~\cite{Billard:2013qya}, and a re-calculation of the neutrino floor is needed for a fair comparison\footnote{The mixture of multiple target nuclei present in NEWSdm will further aid in mitigating against the neutrino background, due to target complementarity, see Ref.~\cite{Ruppin:2014bra}.}. Here, we explore the discovery potential (and impact on the neutrino floor) of time-integrated detectors which have a fixed orientation in the lab frame without an equatorial mount.

\section{Event rate}
\label{sec:eventrate}

\subsection{Formalism}
The directional DM-nucleus scattering rate per unit detector mass as a function of recoil energy $E_r$ and direction $\qhat$ is given by~\cite{Gondolo:2002np},
\begin{equation}\label{eq:directionalrate}
 \frac{\mathrm{d}^2R(t)}{\mathrm{d}E_r \mathrm{d}\Omega_q} = \frac{\rho_0}{4\pi \mu_{\chi p}^2 m_\chi} \sigma_p \mathcal{C}_N F^2(E_r) \hat{f}(\vmin,\qhat;t) \,,
\end{equation}
where $m_\chi$ is the DM mass, $\mu_{\chi p}$ is the DM-proton reduced mass and $\sigma_p$ is the DM-proton cross section for either spin-independent (SI) or spin-dependent (SD) interactions. The function $F^2(E_r)$ is the nuclear form factor parameterising the loss of coherence in the DM-nucleus interaction at high momentum transfer. The coefficient $\mathcal{C}_N$ is an enhancement factor which depends on the nucleon content of the target nucleus $N$, which along with the cross section can encode SI or SD scattering. In this work, we will focus on SI scattering, and equal couplings to protons and neutrons so that $\mathcal{C}_N = A^2$ for a nucleus with $A$ nucleons. We assume the standard Helm ansatz for the form factor~\cite{Helm:1956zz}, which is an excellent approximation for the energy scales of direct detection~\cite{Vietze:2014vsa,Co:2012adt}. 

The DM velocity distribution enters in the form of its Radon transform $\hat{f}$~\cite{Radon,Gondolo:2002np} at $\vmin = \sqrt{m_N E_r / 2\mu_{\chi N}^2}$, the smallest DM speed that can create a recoil of energy $E_r$:
\begin{equation}
\label{eq:Radon}
 \hat{f}(\vmin,\qhat;t) = \int f(\mathbf{v},t) \,\delta(\mathbf{v}\cdot\qhat-\vmin)\, \mathrm{d}^3\mathbf{v} \, .
\end{equation}
We assume the standard halo model (SHM) which yields a Maxwell-Boltzmann velocity distribution (see e.g. Ref.~\cite{Green:2017odb})
\begin{eqnarray}
\label{eq:SHM}
f(\mathbf{v},t) =& \frac{1}{(2\pi \sigma_v^2)^{3/2}N_\mathrm{esc}} \, \exp \left( - \frac{(\mathbf{v} - \mathbf{v}_\textrm{lab}(t))^2}{2\sigma_v^2}\right) \\
&\times \, \Theta (\vesc - |\mathbf{v} - \mathbf{v}_\textrm{lab}(t)|)\,, \nonumber
\end{eqnarray}
with the normalisation constant,
\begin{equation}
N_\mathrm{esc} = \erf \left( \frac{\vesc}{\sqrt{2}\sigma_v}\right) - \sqrt{\frac{2}{\pi}} \frac{\vesc}{\sigma_v} \exp \left( -\frac{\vesc^2}{2\sigma_v^2}   \right)\,.
\end{equation}
The parameters of the SHM velocity distribution are its peak speed $v_0 = 220 \kms$ and width $\sigma_v = v_0/\sqrt{2} \approx 156 \kms$. The velocity distribution is truncated at the escape speed of the Milky Way, for which we use the best fit RAVE measurement $\vesc = 533 \kms$~\cite{Piffl:2013mla}. The observed velocity distribution is found after a Galilean boost into the laboratory frame by $\textbf{v}_\textrm{lab}(t)$. This boost is responsible for time dependence and, for an isotropic Galactic frame velocity distribution, is the sole source of the anisotropy in recoil directions. The lab velocity is the sum of four components,
 \begin{equation}
\textbf{v}_\textrm{lab}(t) = {\bf v}_{\rm GalRot} + {\bf v}_\odot + {\bf v}_{\rm EarthRev}(t) + {\bf v}_{\rm EarthRot}(t) \, ,
\end{equation}
which are the bulk rotation of the local standard of rest (LSR) around the Galactic center, ${\bf v}_{\rm GalRot}$, the peculiar velocity of the Solar System with respect to the LSR, ${\bf v}_\odot$, the Earth's revolution around the Sun, ${\bf v}_{\rm EarthRev}$, and the Earth's rotation, ${\bf v}_{\rm EarthRot}$. In Galactic co-ordinates we set the LSR rotation speed to ${\bf v}_{\rm GalRot} = (0,220,0)$~km~s$^{-1}$, and for the peculiar velocity we use the Schoenrich~et~al.~\cite{Schoenrich:2009bx} determination, $\textbf{v}_\odot = (11.1,12.24,7.25)$~km~s$^{-1}$. The Earth revolution and rotation velocity calculations can be found in, for example, Refs.~\cite{Mayet:2016zxu,McCabe:2013kea}.

\subsection{Detectors}
In this work, we bracket a range of target nuclei masses by considering two benchmarks: xenon and carbon.  Although there are currently no directionally sensitive experiments using xenon\footnote{There have been suggestions that it may be possible to exploit columnar recombination to infer a 1D projection of recoil directions in liquid noble detectors~\cite{Nygren:2013nda,MuAaoz:2014uxa,Li:2015zga,Mohlabeng:2015efa,Nakajima:2015dva,Gehman:2013mra,Cao:2014gns,Cadeddu:2017ebu}.}, its use as an example allows a direct comparison with the results of the most sensitive current detectors (LUX~\cite{Akerib:2016vxi}, PandaX~\cite{Tan:2016zwf} and Xenon1T~\cite{Aprile:2017iyp}), as in previous work~\cite{Grothaus:2014hja,O'Hare:2015mda}. Carbon allows us to explore the dependence of our results on the mass of the target nucleus, and also makes up a significant fraction of the mixture of target nuclei used in NEWSdm.

We assume idealised, background free experiments (apart from neutrino backgrounds in Secs.~\ref{sec:ideal-neutrinos} and~\ref{sec:nonideal-neutrinos}). For simplicity we assume that our mock experiments have perfect electronic/nuclear recoil discrimination. Our results therefore represent an upper limit on the sensitivity of real experiments (however directional experiments are expected in general to have good electronic/nuclear discrimination). We also assume perfect angular resolution, as the impact of realistic resolutions on the directional recoil spectrum are rendered largely unimportant after averaging over many Earth rotations. Furthermore, in order to isolate the effects of time integration, we initially assume that the detector can reconstruct the full three dimensions of a recoil track and measure its sense. In Sec.~\ref{sec:nonideal} we depart from these assumptions.

We assume slightly lower  energy thresholds than are currently achievable: 1 keV for xenon and 40 keV for carbon. Our 1 keV xenon threshold is an optimistic extrapolation of what may be possible beyond the next generation of dual-phase xenon detectors. In the case of carbon, a 40 keV threshold is reasonable and would correspond to a track length threshold of  $\lesssim$100 nm~\cite{Aleksandrov:2016fyr}. This is well above the optical scanning accuracy already demonstrated by NEWSdm ($\sim$10 nm) using X-ray microscopy or a resonant light scattering technique. 
We note that very low recoil energy thresholds will be required in directional detectors if they are to be able to probe cross sections below the neutrino floor, as we will discuss in Secs.~\ref{sec:ideal-neutrinos} and~\ref{sec:nonideal-neutrinos}.  Most results we express in terms of event numbers so that they are independent of the choice of cross section, but we will also give some reference exposures for example cross sections. Generally we consider larger exposures for carbon, since a solid detector will be more easily scalable than gaseous ones.

\subsection{Angular recoil distribution}
\label{sec:distribution}

\begin{figure}[t]
\centering
\includegraphics[width=0.45\textwidth]{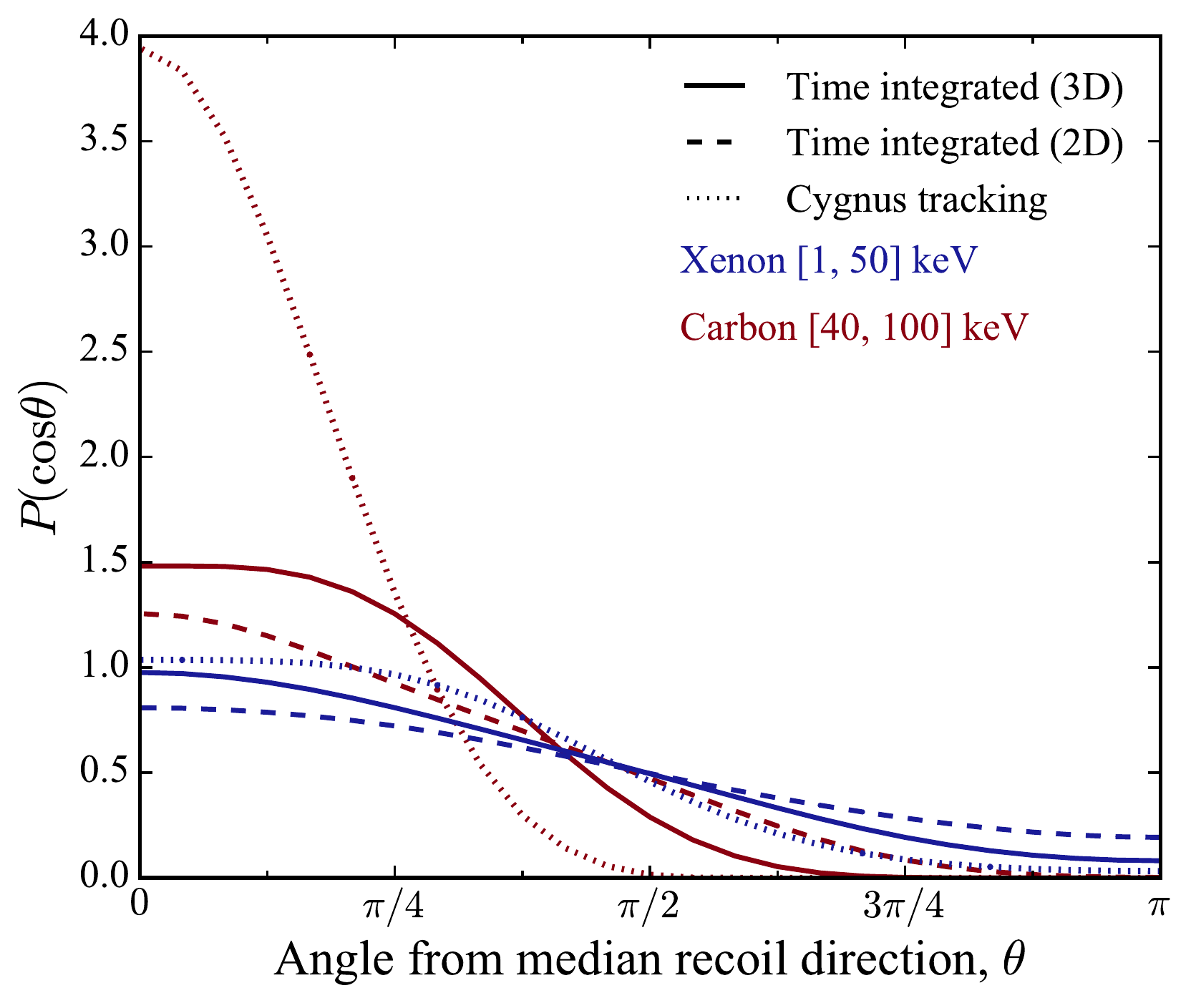}
\caption{Distribution of recoils as a function of $\theta$, the angle from the median recoil direction for experiments with time integration and 3D readout (solid line); time integration and 2D readout (dashed line); and with Cygnus tracking (dotted line).  Results are shown for xenon (blue) and carbon (red) targets for $m_{\chi} = 50 \, {\rm GeV}$. In the 2D readout case, we consider a horizontally-aligned detector at Gran Sasso ($\lambda \approx 42.5^\circ \,\mathrm{N}$).}
\label{fig:Rate}
\end{figure}

The angular distribution of recoils can be obtained from Eqs.~\eqref{eq:Radon}~and~\eqref{eq:SHM}. This distribution exhibits a strong dipole \cite{Spergel:1987kx,Bozorgnia:2011vc}, with the median recoil direction pointing along $-\mathbf{v}_\mathrm{lab}(t)$, assuming the DM velocity distribution is isotropic. This is illustrated in Fig.~\ref{fig:Rate}, where the dotted lines show the expected distribution of the recoil angle $\theta$ (integrated over recoil energies), measured with respect to this median recoil direction, $-\mathbf{v}_\mathrm{lab}(t)$. However, a detector does not measure directions with respect to the lab velocity (we refer to this co-ordinate system as the `Cygnus-tracking' frame). Instead, recoil directions $\qhat$ are measured in a lab-fixed reference frame which we take to have co-ordinate axes along (North, West, Zenith):
\begin{equation}
\label{eq:recoils}
\qhat = \sin \vartheta \cos\varphi \,\hat{\mathcal{N}} + \sin\vartheta \sin\varphi \,\hat{\mathcal{W}} + \cos\vartheta \,\hat{\mathcal{Z}}\,.
\end{equation}
If the time of an event is tagged, however, the event's direction can be transformed back into the Cygnus-tracking frame and the full anisotropy of the signal is preserved.

\begin{figure}[t]
\centering
\includegraphics[width=0.45\textwidth]{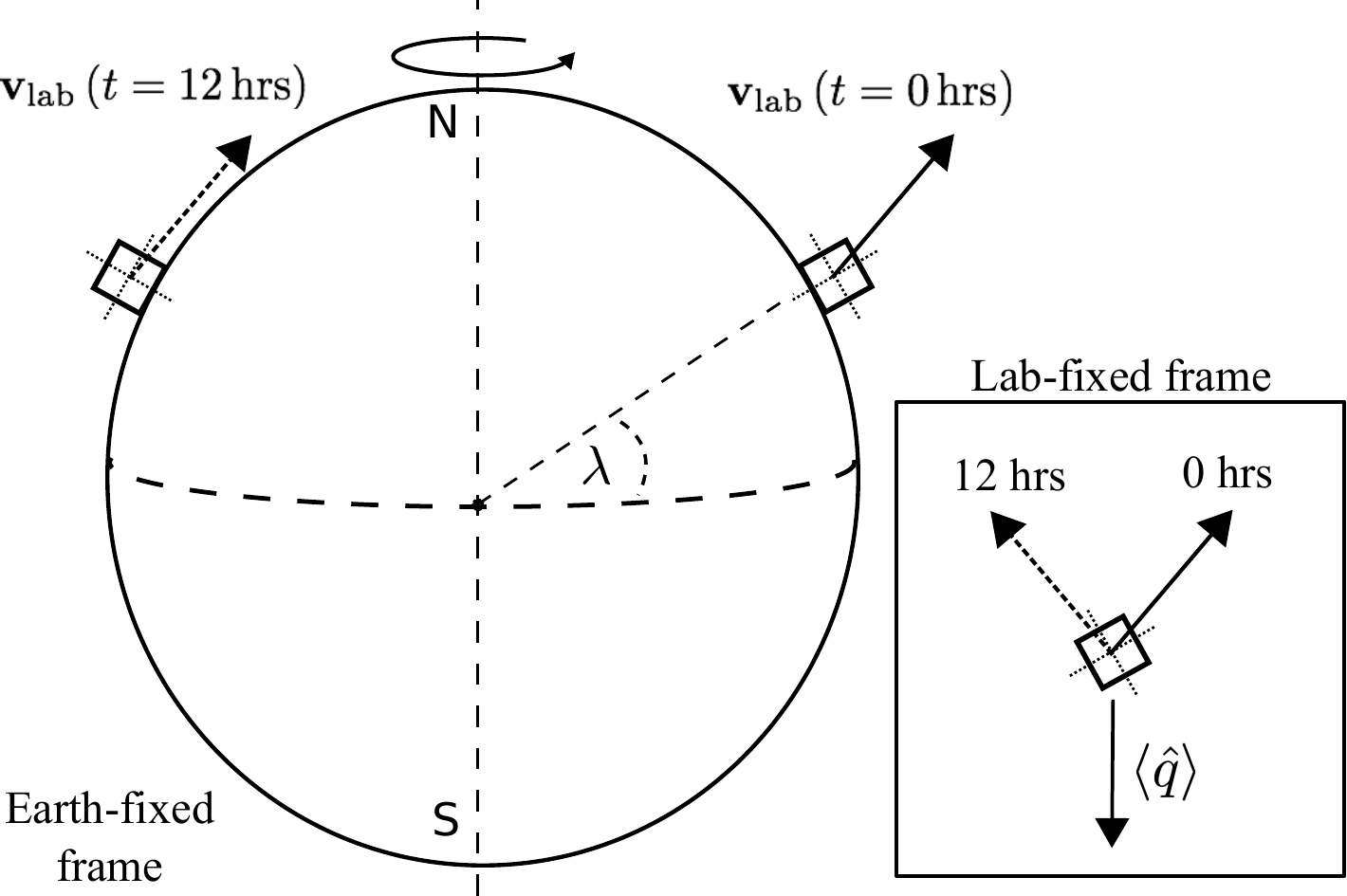}
\caption{Illustration of the lab's velocity $\mathbf{v}_\mathrm{lab}$ in the Earth-fixed and lab-fixed frames. We consider a hypothetical lab at latitude $\lambda$ at two different times of day. The Earth's rotation washes out any asymmetry perpendicular to the rotation axis, with the median recoil direction $\langle \qhat \rangle$  over the course of a day being parallel to the rotation axis. Note that over the course of a year, the angle between $\mathbf{v}_\mathrm{lab}$ and the Earth's rotation axis varies between about $36^\circ$ and $49^\circ$ \cite{Kouvaris:2014lpa}.}
\label{fig:Earth}
\end{figure}

If instead no timing information about individual events is available then a detector can only measure directions in the lab-fixed co-ordinate system of Eq.~\eqref{eq:recoils} and the expected angular distribution of recoils is obtained by integrating Eq.~\eqref{eq:directionalrate} over the total exposure time (which we take to be one year). In Fig.~\ref{fig:Earth}, we sketch the directions of $\mathbf{v}_\mathrm{lab}$ for a hypothetical lab at two times of day, 12 hours apart. Between $t = 0 \,\,\mathrm{hrs}$ and $t = 12 \,\,\mathrm{hrs}$, the lab's velocity and therefore the median recoil direction (as measured in the lab-fixed frame) rotate by $180^\circ$ about the Earth's rotation axis. Summing these two contributions to the recoil distribution, we see that any asymmetry perpendicular to the Earth's rotation axis is washed out. The median recoil direction must then be parallel to the Earth's axis and as such the recoils (which point away from the motion of the lab) must point North-to-South in an Earth-fixed frame, as illustrated in Fig.~\ref{fig:Earth}. In the lab-frame co-ordinate system (North, West, Zenith), this median direction should be:
\begin{align}
\label{eq:mean_recoil}
\left\langle \qhat \right\rangle = -\sin \left(\frac{\pi}{2} + \lambda\right) \, \hat{\mathcal{N}} + \cos \left(\frac{\pi}{2} + \lambda \right) \, \hat{\mathcal{Z}}\,,
\end{align}
for a detector at latitude $\lambda$\footnote{By convention, we take latitudes in the Northern (Southern) hemisphere to be positive (negative).}. We have verified numerically that this is indeed the case.

For a time-integrated detector which can measure the full 3D track direction, we find the angular recoil spectrum to be essentially independent of location since the recoil distributions observed at different latitudes are nearly identical up to a rotation. In Fig.~\ref{fig:Rate} we show this time-integrated angular recoil spectrum (solid lines) as a function of $\theta$, the recoil angle measured from the median recoil direction in Eq.~\eqref{eq:mean_recoil}. As expected the time integration makes the angular distribution less anisotropic. The recoil distributions (both with and without time integration) are more anisotropic for our carbon benchmark experiment. This is because its higher threshold energy means it is only sensitive to higher energy recoiling nuclei which preserve more of the directional preference of the original DM flux.

For detectors with a 2D readout, all that can be measured is the projection of each recoil track onto a 2D plane in the lab-frame. In this case, the observed recoil distributions after time integration will depend upon the orientation of the readout plane as well as the detector latitude. In Fig.~\ref{fig:Rate}, we plot the angular recoil distribution for a 2D detector at Gran Sasso ($\lambda \approx 42.5^\circ \,\mathrm{N}$), with the readout plane oriented horizontally in the (North, West) plane. In the following sections, we consider an experiment with full 3D readout, exploring experiments with 2D readout further in Sec.~\ref{sec:nonideal}.

\section{Ideal detectors}\label{sec:ideal}

\subsection{Rejecting isotropy}\label{sec:ideal-iso}
\begin{figure}[t]
\centering
\includegraphics[width=0.49\textwidth]{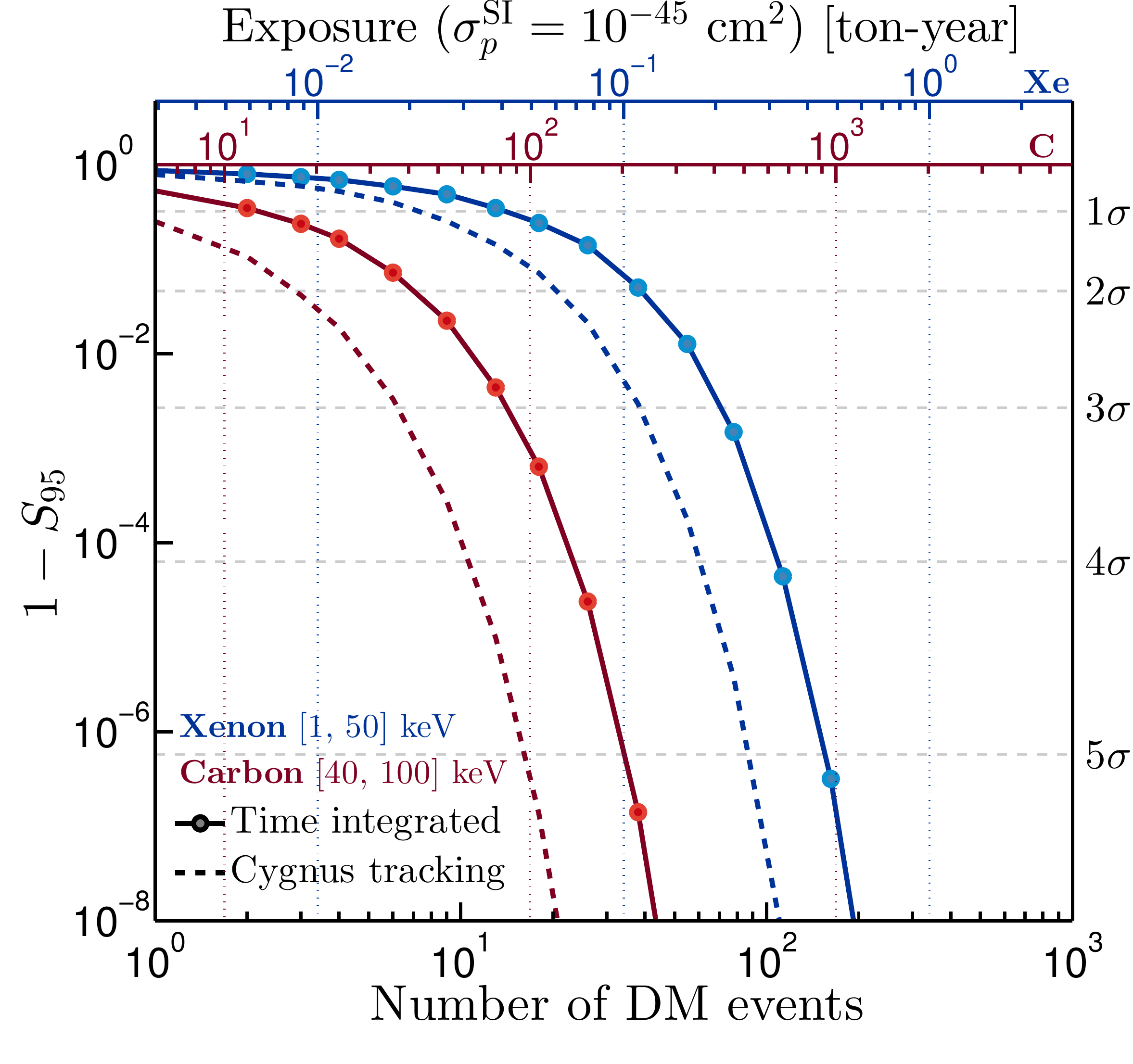}
\caption{The number of DM signal events required to reject isotropy at the $S_{95}$ significance level in 95\% of simulated experiments with (solid line) and without (dashed) time integration for xenon (blue) and carbon (red) for $m_{\chi}~=~50 \, {\rm GeV}$.  The right-hand axis displays the significance in terms of $\sigma$, while the top axes give the exposure in ton-years required for a cross section $\sigma_{p}^{\rm SI} = 10^{-45} \, {\rm cm^2}$ for xenon (blue) and carbon (red).}
\label{fig:Isotropy}
\end{figure}

The first goal of a directional experiment which observes a number of recoil events is to ascertain whether or not their directions are anisotropic. Reference~\cite{Morgan:2004ys} found that the most powerful test for doing this uses the average of the cosine of the recoil directions~\cite{Briggs}:
\begin{equation}
\langle \cos{\theta_\mathrm{rec}} \rangle = \frac{ \sum_{i=1}^{N} \cos{\theta^{i}_\mathrm{rec}}}{N} \,.
\end{equation}
The angles $\theta^i_\mathrm{rec}$ are each measured between the recoil vector $i$ and some chosen preferred direction $\qhat_\mathrm{exp}$. We choose $\qhat_\mathrm{exp} = \qhat_\mathrm{Cyg}$ for a Cygnus-tracking experiment and $\qhat_\mathrm{exp} = \langle \qhat \rangle$, defined in Eq.~(\ref{eq:mean_recoil}), for the time-integrated signal. For isotropic vectors $\langle \cos{\theta_\mathrm{rec}} \rangle$ takes values on the interval $[-1, 1]$ with a Gaussian distribution with mean 0 and variance $(1/3N)$. For DM-induced recoils, the distribution of $\cos\theta_\mathrm{rec}$ is instead as shown in Fig.~\ref{fig:Rate}, with a different distribution for Cygnus-tracking and time-integrated experiments.

For each set of inputs we build distributions from $10^4$ Monte Carlo experiments, assuming that the observed recoil directions are distributed in the same way as DM-induced recoils. Our statistical analysis consists of a frequentist hypothesis test on the simulated distributions. We take as the null hypothesis that the recoil directions are isotropic. 
This allows us to calculate the significance with which the isotropic case can be rejected for a given measured value of $\langle \cos{\theta_{\mathrm{rec}}} \rangle$, based on the cumulative null distribution. We then quantify the success of the test in terms of $S_{95}$ which we define as the minimum significance level achievable in 95\% of simulated experiments. In practice the expected preferred direction does not necessarily need to be known a priori to perform the test. The null distribution for isotropic vectors is recovered independent of the choice of $\qhat_\mathrm{exp}$, but for anisotropic recoils the significance of a given result is maximised by choosing the correct preferred direction.
 
In Fig.~\ref{fig:Isotropy}, we show the number of events required to reject isotropy as a function of the significance level, $S_{95}$ using the $\langle \cos\theta_\mathrm{rec} \rangle$ statistic on both benchmark targets for experiments with 3D track reconstruction which integrate over time (solid) or track Cygnus (dashed). The DM mass assumed here was $m_\chi = 50$~GeV. We also give the exposure required to accumulate this number of events for a cross section 
$\sigma_{p}^{\rm SI} = 10^{-45} \, {\rm cm^2}$. Since the number of events is directly proportional to the product of the exposure and cross section the scaling to other cross sections is straightforward.

Time integration reduces the significance for rejecting isotropy, as expected since the time-integrated recoil distribution is   more isotropic, as shown in Fig.~\ref{fig:Rate}. For a given significance, roughly twice as many events are required for a time-integrated experiment to reject isotropy, compared with a Cygnus-tracking experiment. Comparing results between the two target nuclei: the lower threshold xenon experiment requires roughly 5 times more events to reject isotropy than the higher threshold carbon experiment for both the time-integrated and Cygnus-tracking cases. This reflects the fact that the high energy recoils are more anisotropic, and hence fewer events are required to establish that the distribution is not isotropic. The rate of high energy recoils is substantially lower and the exposure necessary to accumulate the required number of events is roughly two orders of magnitude larger for the high-threshold carbon experiment than for the low-threshold xenon experiment. We emphasise though that a solid experiment can be more easily scaled to large exposures than a gaseous one. This target/threshold dependence in the results from model independent tests would likely not be present to the same degree in a likelihood-based approach using the expected recoil energy and direction distributions~\cite{Billard:2009mf,Billard:2010jh}. However the analysis methodology we adopt here has the advantage of not requiring a parameterisation of the recoil distribution, as would be needed to construct a likelihood.

\subsection{Measuring median recoil direction}\label{sec:ideal-med}
\begin{figure*}[t!]
\centering
\includegraphics[width=0.49\textwidth]{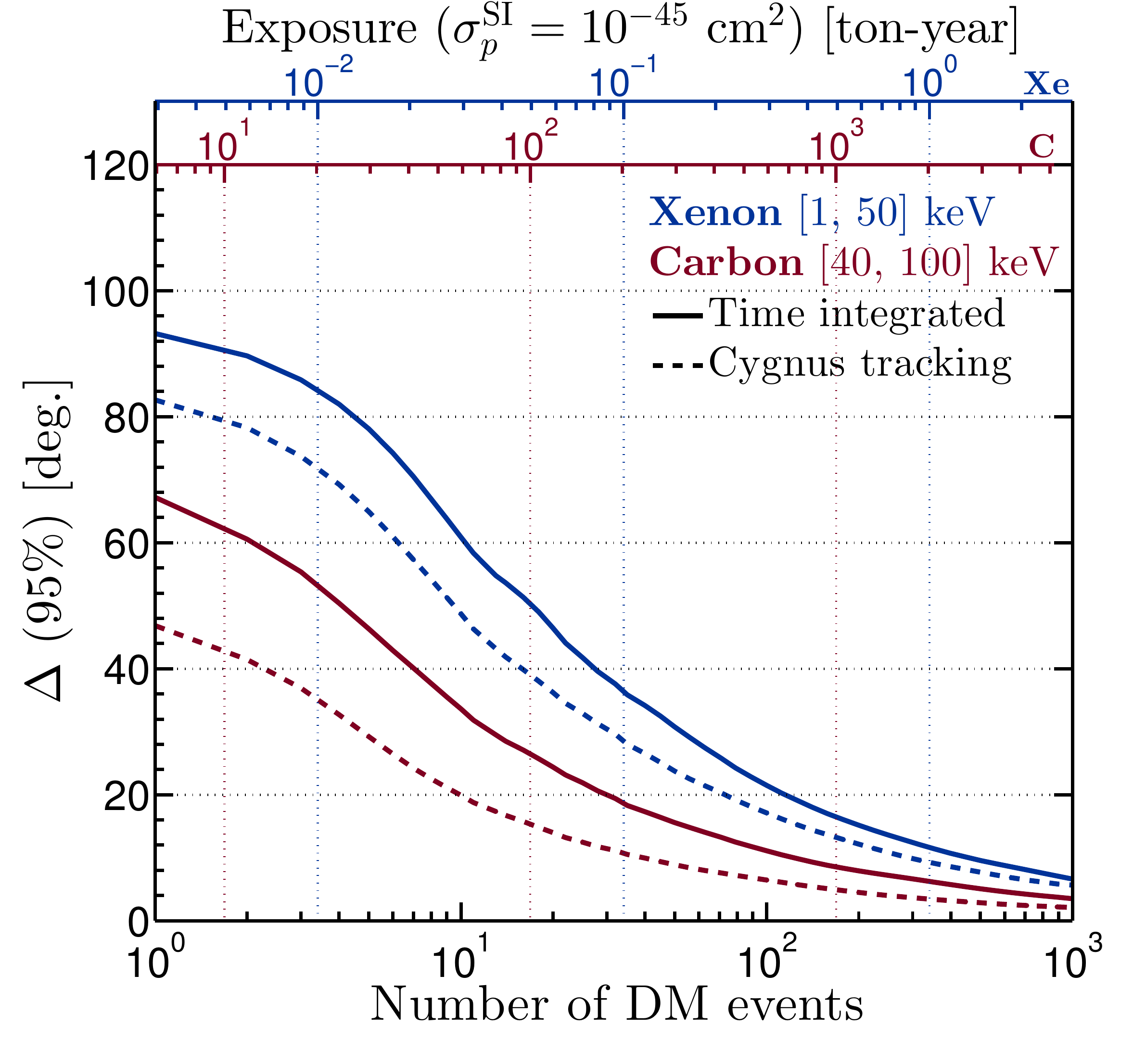}
\includegraphics[width=0.49\textwidth]{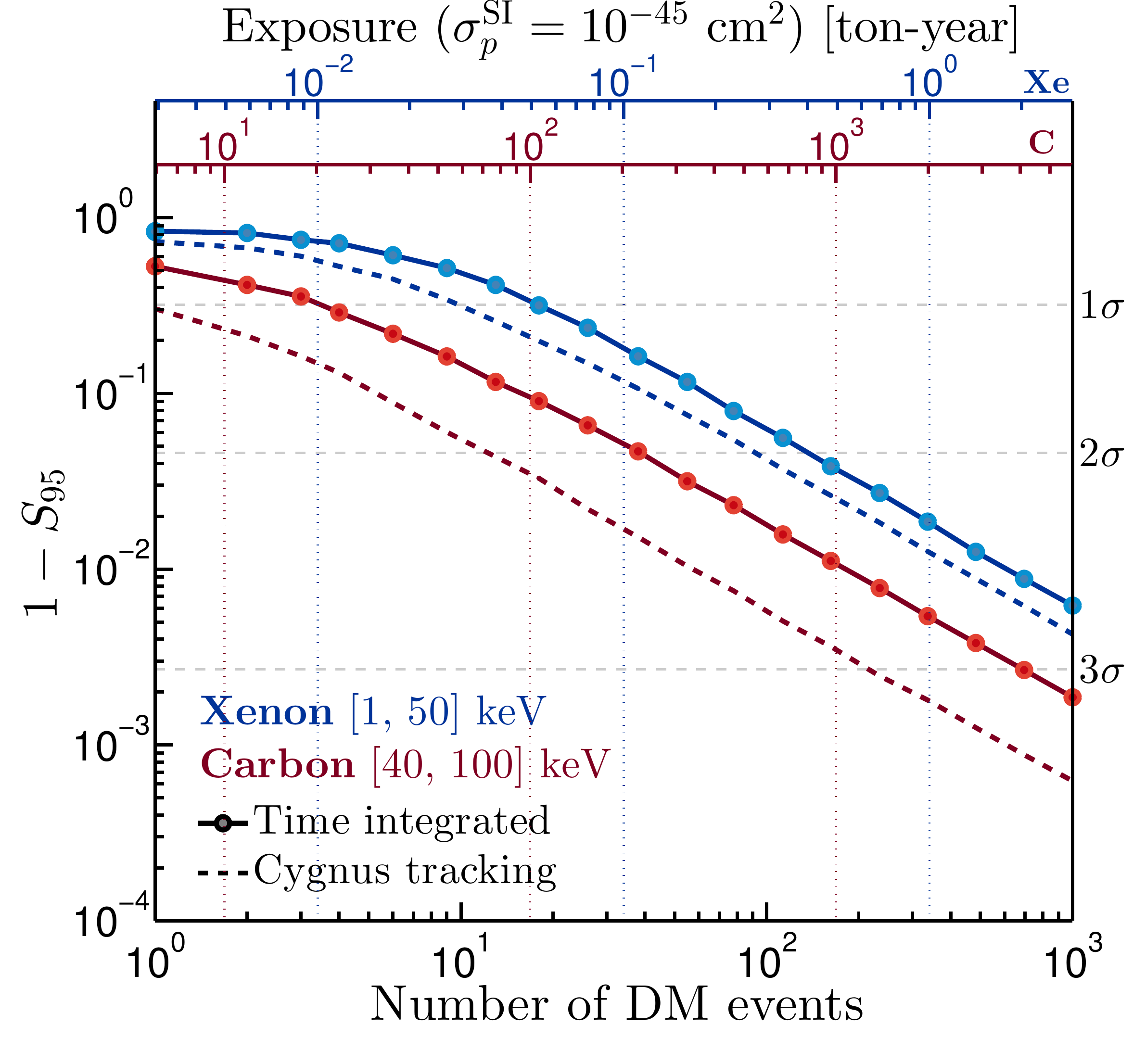}
\caption{{\bf Left:} The 95th percentile of $\Delta$, the angle between the observed median recoil direction and the expected signal median recoil direction (defined in Eq.~(\ref{eq:Delta})), as a function of the number of DM events (circles removed from solid line for clarity).  {\bf Right:} The number of DM events required to reject a random median recoil direction at a given level, $S_{95}$, in 95\% of simulated experiments. The right-hand axis gives this significance in terms of $\sigma$. Line types are as in Fig.~\ref{fig:Isotropy}.
}
\label{fig:MedianDirection}
\end{figure*}
Once it has been established that a set of measured recoils are anisotropic, the next step is to measure the median recoil direction and confirm that it matches the expectation for DM-induced recoils. As discussed in Sec.~\ref{sec:distribution}, for an isotropic DM velocity distribution the expected median recoil direction coincides with the inverse of the direction of Solar motion for a Cygnus-tracking experiment, while for a time-integrated experiment the median recoil direction is parallel to the Earth's rotation axis, Eq.~\eqref{eq:mean_recoil}.  We follow the method of Ref.~\cite{Green:2010zm} and use a test statistic based on $\Delta$, the angle between the inverse median recoil direction $\qhat_\mathrm{med}$ and the expected median direction $\qhat_\mathrm{exp}$~\cite{Fisher}: 
\begin{align}\label{eq:Delta}
\Delta = \cos^{-1}(\qhat_\mathrm{med}\cdot\qhat_\mathrm{exp})\,.
\end{align}

The median direction is defined as the direction which minimises the sum of the arclengths between itself and the individual inverse recoil directions. Isotropically distributed recoil vectors have no expected median direction, so the statistic $(1 - \cos\Delta)/2$ will follow a uniform distribution in the range $[0,1]$. This forms the null distribution we use to extract a significance value for a given measured $\Delta$. As before we use the Monte Carlo generated distribution of $\Delta$ assuming DM-induced recoils to extract $S_{95}$, the minimum significance level achievable by 95\% of simulated experiments.

The left-hand panel of Fig.~\ref{fig:MedianDirection} shows the 95th percentile of the angle between the median inverse recoil direction and the direction of Solar motion, $\Delta$, as a function of the number of DM events. Again, the DM mass was taken to be $m_\chi = 50$~GeV. We find that the xenon (carbon) benchmark experiment requires a factor of 1.6 (3) times more events to measure the median recoil direction to within 20$^\circ$ of the true median in 95\% of experiments if they are time-integrating rather than Cygnus-tracking. The right-hand panel shows how many events are needed to reject a randomly oriented median recoil direction. 
As in Fig.~\ref{fig:Isotropy}, we also show the exposure required to accumulate this number of events for both benchmark targets for a cross section 
$\sigma_{p}^{\rm SI} = 10^{-45} \, {\rm cm^2}$.  The number of signal events required to achieve a 3$\sigma$ agreement with the expected median direction in 95\% of experiments is increased by a factor of 1.4 (3) for a time-integrating xenon (carbon) experiment. This factor is larger for the higher threshold carbon experiment because its signal loses a greater degree of anisotropy after time integration, as can be seen in Fig.~\ref{fig:Rate}.

\subsection{The neutrino floor}\label{sec:ideal-neutrinos}
\begin{figure*}[t]
\centering
\includegraphics[width=0.98\textwidth]{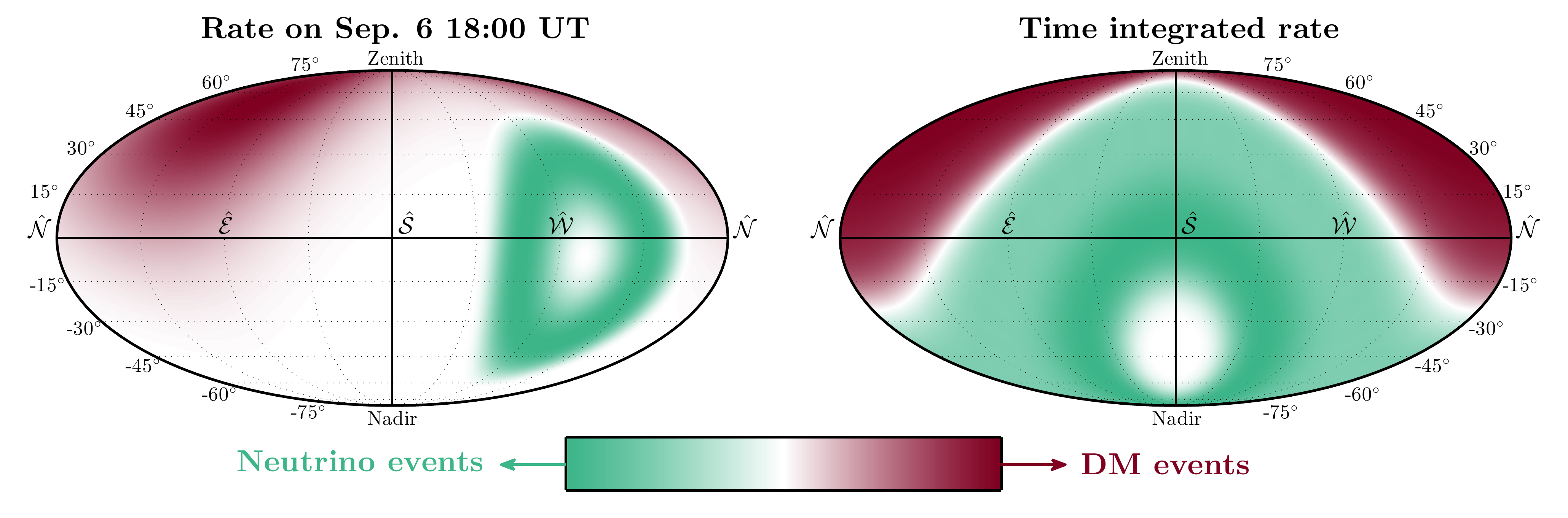}
\caption{Mollweide projection of the xenon angular event rate at Gran Sasso, integrated over the energy window of the detector [1, 50] keV, for a 6 GeV DM particle. The projection is in the laboratory co-ordinate system with South located at the center of each map. Green indicates the rate of $^8$B neutrino events whereas red indicates the rate of DM events. {\bf Left:} Rate on the 6th September at 18:00 {\bf Right:} Time-integrated event rate over 100 days (from Jan 1).  }
\label{fig:skymaps}
\end{figure*}

\begin{figure}
\centering
\includegraphics[width=0.49\textwidth]{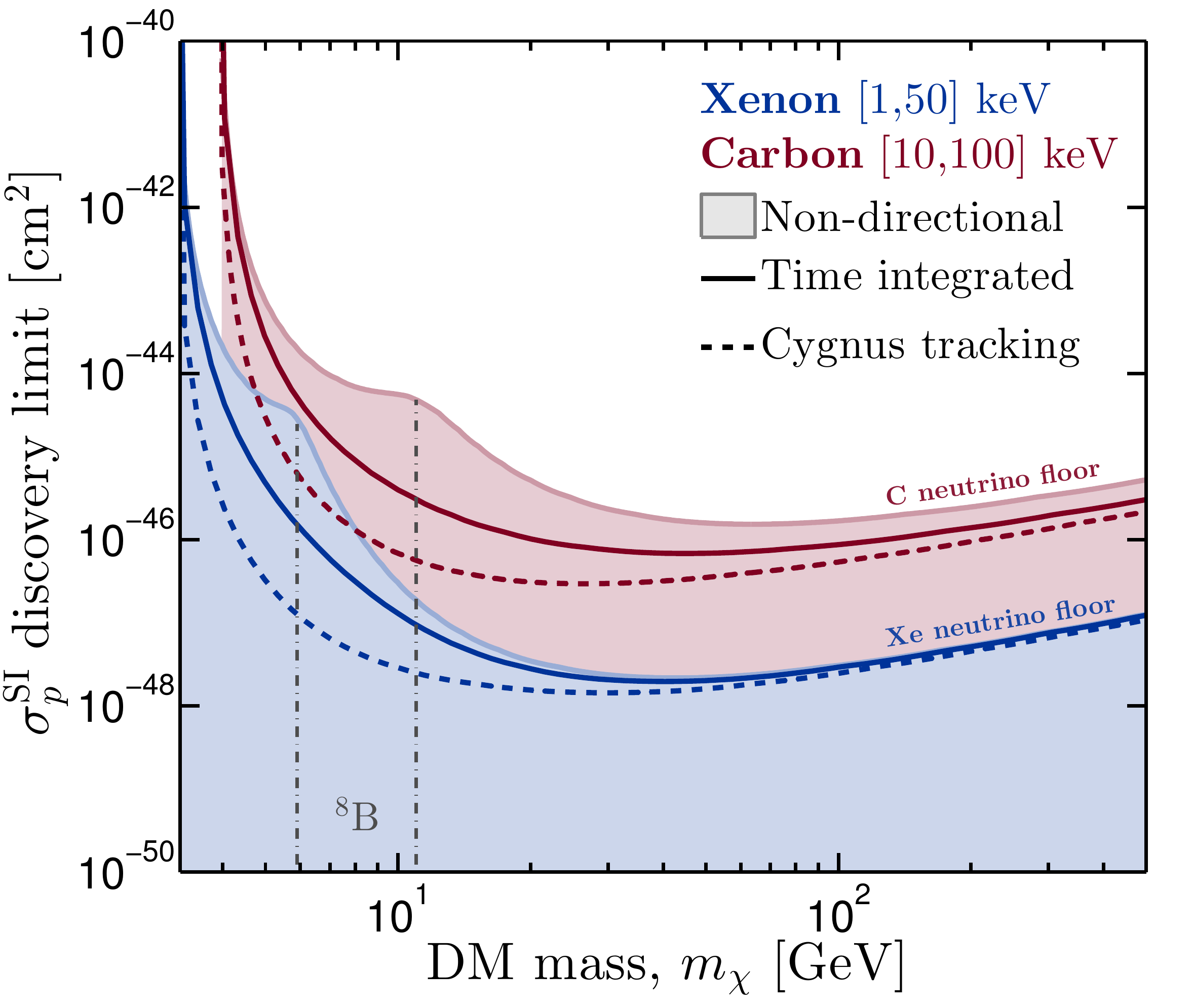}
\caption{The spin-independent neutrino floor and discovery limits as a function of DM mass for a xenon experiment with a [1,~50]~keV energy window and a  10 ton-year exposure (blue) and a carbon experiment with a [10,~100]~keV energy window and a 100 ton-year exposure (red). The dotted (solid) lines are for a Cygnus-tracking (time-integrated) detector and the shaded regions are for a standard non-directional experiment. The vertical dot-dashed lines show the DM masses chosen for Fig.~\ref{fig:NuFloor_DetectorMass}, where the DM recoil energy spectrum most closely matches that of the  $^8$B neutrinos: 6 GeV for xenon and 11 GeV for carbon.}
\label{fig:NuFloor_WIMPMass}
\end{figure}

\begin{figure}
\centering
\includegraphics[width=0.49\textwidth]{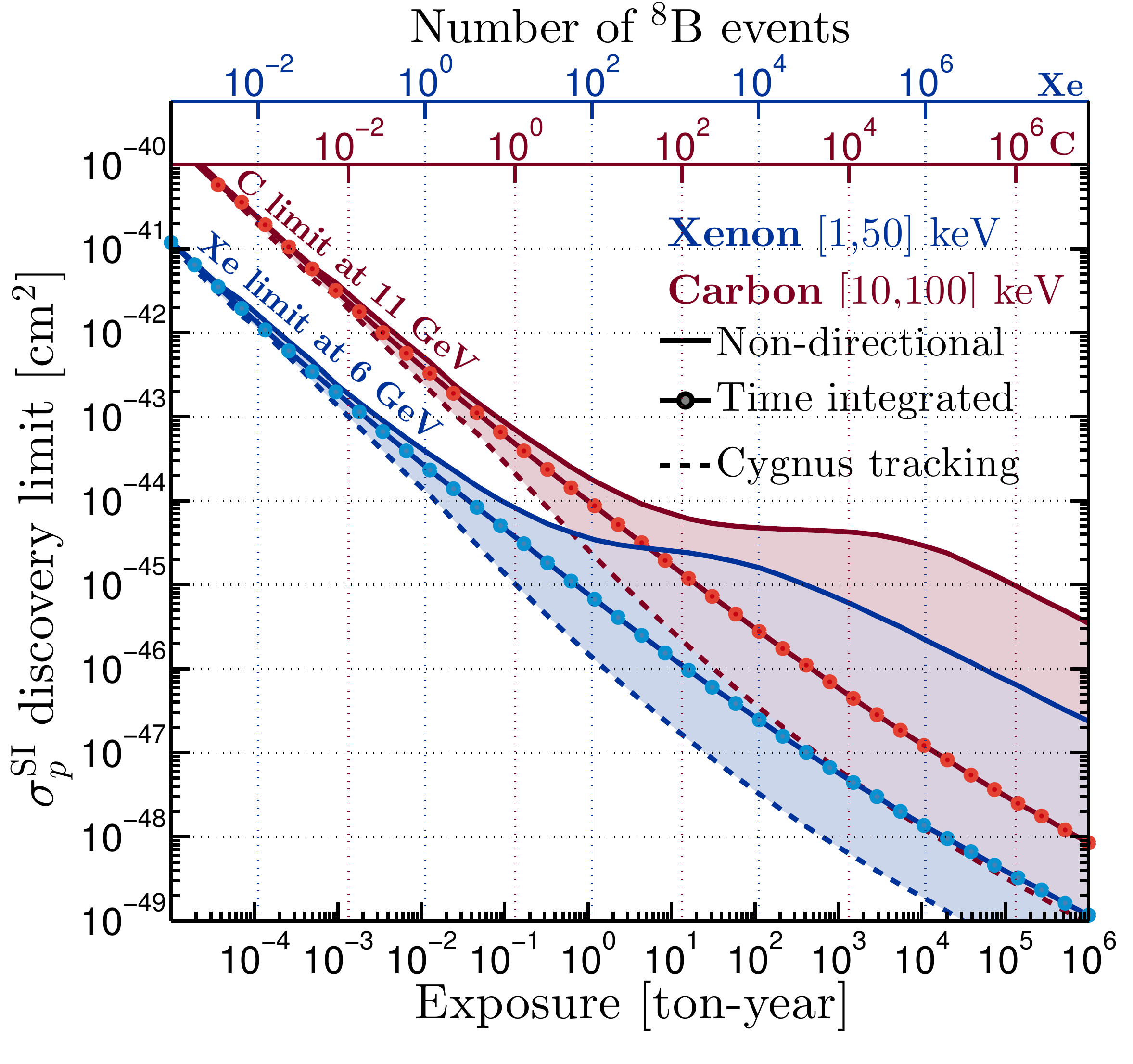}
\caption{The evolution of discovery limits in the presence of neutrino backgrounds as a function of detector exposure at fixed DM masses:  6 GeV for xenon (blue) and 11 GeV for carbon (red), where the DM recoil energy spectrum most closely matches that of the $^8$B neutrinos.
The dotted (solid with circles) lines are for a Cygnus-tracking (time-integrated) detector and the shaded regions are for a standard non-directional experiment. The top axes show the number of $^8$B neutrino events for each target.}
\label{fig:NuFloor_DetectorMass}
\end{figure}
Neutrino backgrounds will limit the sensitivity of conventional non-directional direct detection experiments in the near future. Boron-8 ($^8$B) Solar neutrinos produce nuclear recoils with a similar energy spectrum to a light DM particle (the exact mass depending on the target nucleus) while diffuse supernova neutrino background (DSNB) and atmospheric neutrinos mimic heavier masses~\cite{Strigari:2009bq}. Since the threshold of a NEWSdm-like detector is too high to detect very low energy Solar neutrinos ($^7$Be, $pp$, $pep$ and CNO) we will focus here on the highest energy Solar neutrinos ($^8$B and $hep$). To be consistent with previous calculations of the neutrino floor we normalise our Solar neutrino fluxes to the high-metallicity Solar model~\cite{Serenelli:2011py}.  Above energies of $E_\nu \sim 20$ MeV the neutrino flux becomes dominated by the diffuse background of cosmological supernova neutrinos before being overtaken at 40 MeV by the low energy tail of atmospheric neutrinos from cosmic ray collisions in the upper atmosphere. We use DSNB and atmospheric neutrino fluxes from Refs.~\cite{Beacom:2010kk} and~\cite{Honda:2011nf} respectively, setting the recommended conservative uncertainties of 50\% and 20\%. These neutrinos induce the (xenon) floor above masses of 20 GeV and below SI DM-proton cross sections of $10^{-48}$~cm$^2$. Both the DSNB and atmospheric neutrinos require detector exposures in excess of 1000 ton-years to observe an appreciable rate, so these will comprise a subdominant background for a NEWSdm-like detector.  We calculate the directional neutrino flux as outlined in Ref.~\cite{O'Hare:2015mda}. Key details are given in Appendix~\ref{nu}.

In Fig.~\ref{fig:skymaps} we compare the (energy-averaged) angular part of the recoil spectra for a 6 GeV DM particle and $^8$B neutrinos scattering on a xenon target measured on September 6th at Gran Sasso (when the separation between the distributions is largest) with that integrated over 100 days. The angular rate is displayed using a Mollweide projection onto a 2D plane. Following the convention of `WIMP astronomy' we display the distribution of the inverse recoil vectors, i.e. we show the point on the sky from which the recoils appear to originate, rather than the angles defining $\qhat$. We see that even in the time-integrated case the DM and neutrino recoils occupy different parts of the sky, implying that the neutrino background should still be subtractable even if events are not time-tagged.

Following the existing literature on the neutrino background to direct detection (see e.g.\ Refs.~\cite{Billard:2013qya,Ruppin:2014bra,O'Hare:2015mda,OHare:2016pjy,Dent:2016iht,Franarin:2016ppr,Bertuzzo:2017tuf}), we define the neutrino floor as a discovery limit delimiting cross sections for which 90\% of hypothetical experiments can achieve a $3\sigma$ detection of DM. This limit is computed with a profile likelihood analysis using a DM+neutrino background likelihood with only the DM mass, cross section and neutrino flux normalisations as free parameters (see e.g.\ Ref.~\cite{Billard:2013qya} for details). These results are a best case scenario as we assume perfect energy resolution, electronic/nuclear recoil discrimination, no astrophysical uncertainties, and negligible backgrounds except for neutrinos. As such we can isolate the advantage of a time-integrated directional experiment over a non-directional version of the same detector. 

Since the neutrino background is mimicked by different DM parameters for different target nuclei~\cite{Ruppin:2014bra}, the neutrino floors for xenon and carbon must be calculated separately. To allow our carbon experiment to experience an appreciable neutrino background we now reduce the threshold to 10 keV. Because of the low event rate for coherent neutrino-nucleus scattering we must also simulate mock experiments with exposures exceeding those of the current generation of dark matter experiment. In Fig.~\ref{fig:NuFloor_WIMPMass}, we show the neutrino floor as a function of DM mass for both experiments for a fixed detector mass and an exposure time of 1 year. The xenon limits are calculated for a 10 ton-year exposure whereas the carbon limit is for a 100 ton-year experiment.   We see that the discovery limits for a Cygnus-tracking detector are only slightly better than those for a detector which integrates over time. This is because, as shown in Fig.~\ref{fig:skymaps}, the time-integrated neutrino and DM recoil angular distributions are still significantly different.  In both cases the discovery limits are significantly better than those from a standard non-directional experiment for $m_{\chi} \sim {\cal O} (1-10)$~GeV.

In Fig.~\ref{fig:NuFloor_DetectorMass} we show the evolution of the discovery limit at fixed DM masses as a function of the exposure for the same two experiments. We pick DM masses where the DM recoil spectrum coincides most closely with that of the $^8$B neutrinos: 11 GeV for carbon and 6 GeV for xenon. As previously found (e.g.~Ref.~\cite{Billard:2013qya}), once the expected number of $^8$B neutrino events becomes of order unity the discovery limit for a non-directional experiment plateaus (until the number of events becomes sufficiently large that small differences in the energy spectra can be resolved). The plateauing of the discovery limit that extends over several orders of magnitude in exposure is the principal limitation caused by the neutrino background. We find here that this `floor' is conquered with a time-integrated directional detector, as indeed it is with a fully time-resolved directional detector, in agreement with the results of Ref.~\cite{O'Hare:2015mda}. The Cygnus-tracking limits represent the best possible discovery limit evolution. In this case the discovery limit decreases proportional to $1/N$ (where $N$ is the total number of background events) until around $N\sim10^4$ at which point a Poissonian subtraction $1/\sqrt{N}$ regime takes over. The exposure at which this scaling begins is controlled by the degree of overlap between the signal and background. This is why the time-integrated limits suffer a small loss in sensitivity, although the limits are still a significant improvement over the non-directional case. We find that a factor of $7 - 10$ larger exposure is required to reach a cross section an order of magnitude below the `floor'.

\section{Non-ideal detectors}\label{sec:nonideal}
\begin{figure}
\centering
\includegraphics[width=0.48\textwidth]{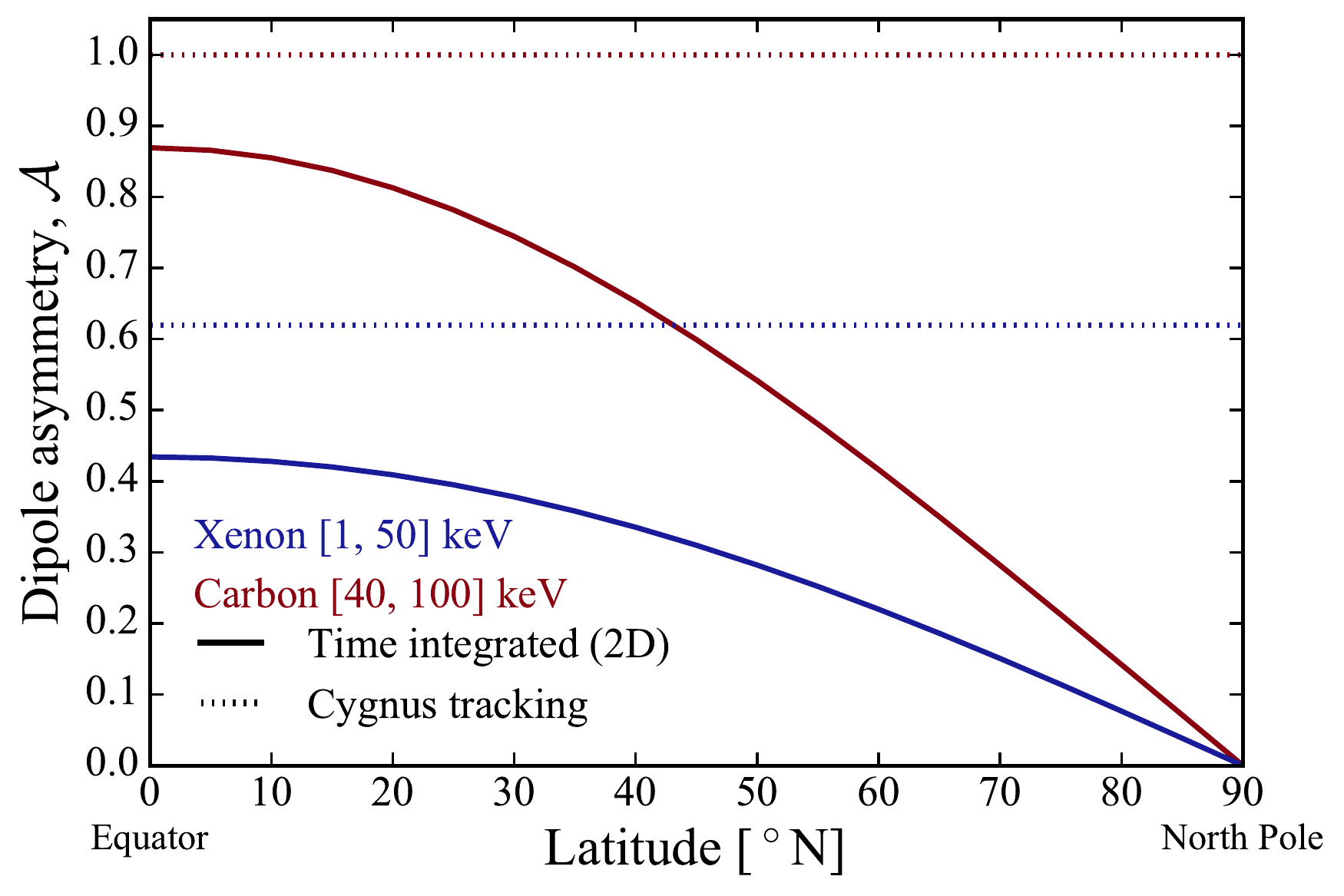}
\caption{Recoil dipole asymmetry $\mathcal{A}$, defined in Eq.~\eqref{eq:asymmetry}, as a function of detector latitude. Dotted lines correspond to Cygnus-tracking detectors, while solid lines correspond to detectors with 2D readouts. Results are shown for xenon (blue) and carbon (red) targets for $m_{\chi} = 50 \, {\rm GeV}$. We show only Northern latitudes; $\mathcal{A}$ is symmetric about the equator.}
\label{fig:asymmetry}
\end{figure}
In order to isolate the effects of time integration, the results in the previous section assume idealised directional detectors, with 3D track readout and sense measurement. However when assessing the merits of time integration versus Cygnus tracking it is crucial to assess whether the benefit of directionality is preserved in realistic experiments that may have limitations in their reconstruction of the recoil track directions. In particular, two experimental limitations are pertinent for a NEWSdm-like experiment using nuclear emulsions, namely 2D track readout and a lack of sense recognition. The ultimate nuclear emulsion detector would in principle exploit multiple layered plates to reconstruct the component of the recoils perpendicular to the plane of the emulsion plates. However currently only readout of the projection onto the emulsion plate has been demonstrated. Similarly, the measurement of the forward-backward sense of each track would require a head-tail effect (i.e.~an asymmetry along the track which allows the beginning to be distinguished from the end~\cite{Majewski:2009an}) for recoils in the emulsion, and this has not been observed to date. Furthermore these limitations have been shown in other cases to severely inhibit the discovery reach of an experiment~\cite{Green:2010zm,Billard:2014ewa}. We repeat the analysis of Secs.~\ref{sec:ideal-iso}--\ref{sec:ideal-neutrinos} for identical mock experiments, but removing the ability to (1) measure the zenithal component of each recoil track and (2) measure the forward or backward going sense of each track. 

In the case of 2D detectors, if we assume that the experiment is oriented such that the plates are parallel to the floor, we also expect the discovery reach to depend on the latitude of the detector. This is because the anisotropy remaining in the 2D time-integrated signal will be washed out to a greater or lesser extent depending on the orientation of the detection plane with respect to the rotation axis of the Earth. For instance, if an experiment were located on the North pole then the rotation of the Earth would wash out all the anisotropy since the plane of detection is parallel to the rotation. However, if a detector were located on the equator then the rotation only washes out the anisotropy in the vertical direction, which is not measured. This effect can be quantified by calculating
\begin{equation}
\label{eq:asymmetry}
\mathcal{A} = \frac{(N_\mathrm{forward} - N_\mathrm{backward})}{(N_\mathrm{forward} + N_\mathrm{backward})}\,,
\end{equation}
where $N_\mathrm{forward}$ is the number of recoil events in the same hemisphere as the mean recoil direction (i.e.~with $\theta \in [0, \frac{\pi}{2}]$) and $N_\mathrm{backward}$ is the number of events in the opposing hemisphere. In a detector with only a 2D readout, the mean recoil direction will be the projection of Eq.~\eqref{eq:mean_recoil} onto the horizontal (North, West) plane. With no zenithal component the mean recoil will always point towards the South, regardless of the latitude of the detector. 

In Fig.~\ref{fig:asymmetry}, we plot the recoil asymmetry $\mathcal{A}$ as a function of latitude, noting that Southern latitudes show the same behaviour as Northern latitudes. Dotted lines show the asymmetry for Cygnus-tracking detectors, while solid lines show the asymmetry for time-integrated detectors with 2D readouts in the (North, West) plane. At large latitudes, the mean recoil direction is roughly perpendicular to the readout plane and so a 2D readout will detect very little asymmetry in the recoil directions. At the equator, the mean recoil direction lies in the readout plane and so the asymmetry is maximised. 

\subsection{Rejecting isotropy}\label{sec:nonideal-iso}
\begin{figure}
\centering
\includegraphics[width=0.49\textwidth]{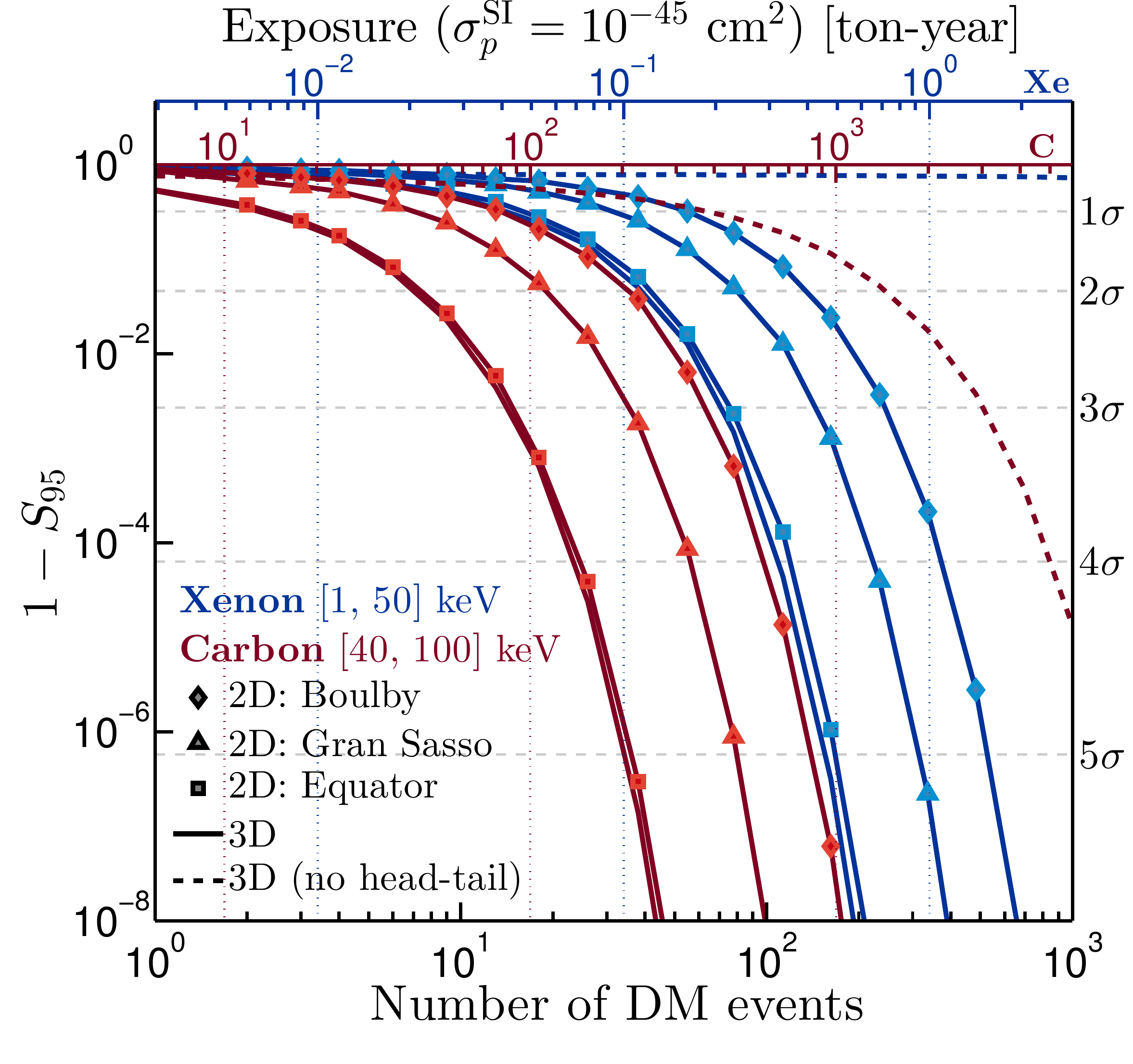}
\caption{The number of DM events required to reject isotropy at the $S_{95}$ level in 95\% of simulated experiments for $m_{\chi}~=~50 \, {\rm GeV}$. The markers show the number of events required in a 2D experiment located at Boulby ($\blacklozenge$), Gran Sasso ($\blacktriangle$) and the equator ($\filledmedsquare$). The solid and dashed lines without markers are for full 3D experiments, but in the latter case head-tail sensitivity has been removed. All curves are for  experiments that are time-integrated and do not track Cygnus.}
\label{fig:Isotropy2D}
\end{figure}
As in Fig.~\ref{fig:Isotropy}, we repeat the analysis for the test of isotropy but now including 2D readout experiments and experiments without head-tail sensitivity. For 2D recoils the $\langle \cos{\theta_{\rm rec}}\rangle$ statistic for isotropic vectors follows again a Gaussian distribution centered on zero, but now with a variance of $1/2N$ rather than $1/3N$. To account for the lack of sense recognition we use the same test but measure the absolute value, $|\langle\cos{\theta_{\rm rec}}\rangle|$. Again the null distribution under isotropic vectors follows a Gaussian distribution with variance of $1/3N$ but instead centered on 0.5. The results of this test are shown in Fig.~\ref{fig:Isotropy2D}. To show the location dependence of the 2D experiments we choose three detector latitudes: the equator (0$^\circ$ N), Gran Sasso (42.5$^\circ$ N) and Boulby (54.6$^\circ$ N). 

The equator is the optimum location and, as expected, the results for a 2D readout at the equator are very close to those with a full 3D readout. Going to Gran Sasso and Boulby at higher latitudes (where there are existing laboratories containing dark matter experiments) makes the 2D signal less anisotropic and therefore increases the number of events required to reject isotropy. Compared to the 3D time-integrated case, roughly twice as many events are required for a 2D readout at Gran Sasso and four times as many events are required at Boulby. 

With no head-tail discrimination (dashed lines), it becomes much more difficult to reject isotropy, requiring around 500 events in a high-threshold carbon experiment to achieve at least a $3\sigma$ rejection in 95\% of experiments. For a low-threshold xenon target without head-tail discrimination, a huge number of signal events ($\gg 10^{3}$) would be required. As shown in Fig.~\ref{fig:Rate}, the angular distribution of recoils is less anisotropic in our benchmark xenon experiment than in the carbon benchmark and without head-tail discrimination this anisotropy is reduced even further. 
We note that even for a Cygnus-tracking experiment a lack of head-tail discrimination would significantly increase the number of events required.

\subsection{Measuring median direction}\label{sec:nonideal-med}
\begin{figure}
\centering
\includegraphics[width=0.49\textwidth]{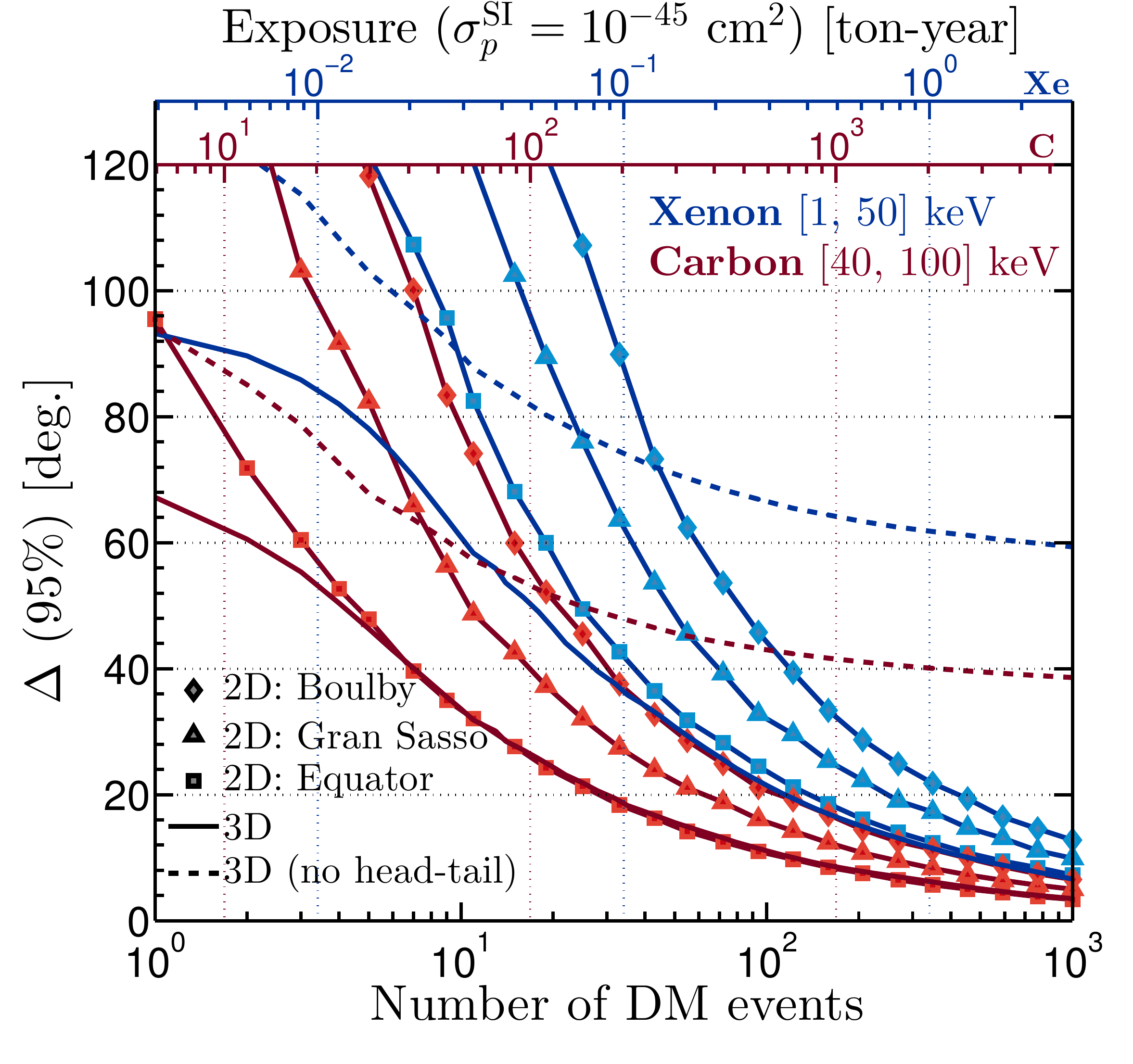}
\caption{All lines as in Fig.~\ref{fig:Isotropy2D} but now displaying the 95th percentile of $\Delta$, the angle between the median inverse recoil direction and the direction of Solar motion.}
\label{fig:MedianDirection2D}
\end{figure}

In Fig.~\ref{fig:MedianDirection2D} we show the 95th percentile values of the angle between the sample median and the expected DM recoil median direction. We again repeat the analysis as presented in Fig.~\ref{fig:MedianDirection}. The median $\phi$ angle describing two-dimensional unit vectors is,
\begin{equation}
 \phi_{\rm med} = \tan^{-1}\left(\frac{\sum_i \sin \phi_i}{\sum_i \cos \phi_i}\right) \, .
\end{equation}
We find the 2D detectors struggle to confirm the correct median direction to within 120$^\circ$, especially with lower event numbers. This is the case for $N<10$ in the high-threshold carbon experiment and $N<40$ for the lower-threshold xenon experiment (which has a greater number of events scattering with large angles from the median). However as the number of observed events increases our 2D results converge on the previous results for a 3D detector. Particularly in the case of a carbon experiment located at the equator, with more than 6 events there is essentially no loss in its capabilities in 2D. For experiments located at our highest latitude (Boulby) we find that measuring the median direction to within 20$^\circ$ of the true underlying median a 2D detector would require a factor of 3.7 times more signal events in both xenon and carbon experiments.

As in the previous subsection we again find the crucial need for head-tail recognition in a directional detector. Using recoils without sense information, the median direction can only be measured down to $\sim 60^\circ\,(\sim 40^\circ)$ in the xenon (carbon) experiment even for event numbers in excess of 1000. Interestingly we note that for very small event numbers, the 3D detector without head-tail outperforms the corresponding 2D detector with head-tail at some locations, but the 95th percentile for $\Delta$ is still in excess of 90$^\circ$ with sample sizes this small. However, we note that if a 2D nuclear emulsion experiment could be oriented at a fixed angle, in order to mitigate against its latitude on Earth and maintain the median recoil direction in the 2D readout plane, then it will be much less important to achieve 3D readout. We discuss this idea further in Sec.~\ref{sec:conclusions}.

\subsection{The neutrino floor}\label{sec:nonideal-neutrinos}
\begin{figure*}
\centering
\includegraphics[width=0.49\textwidth]{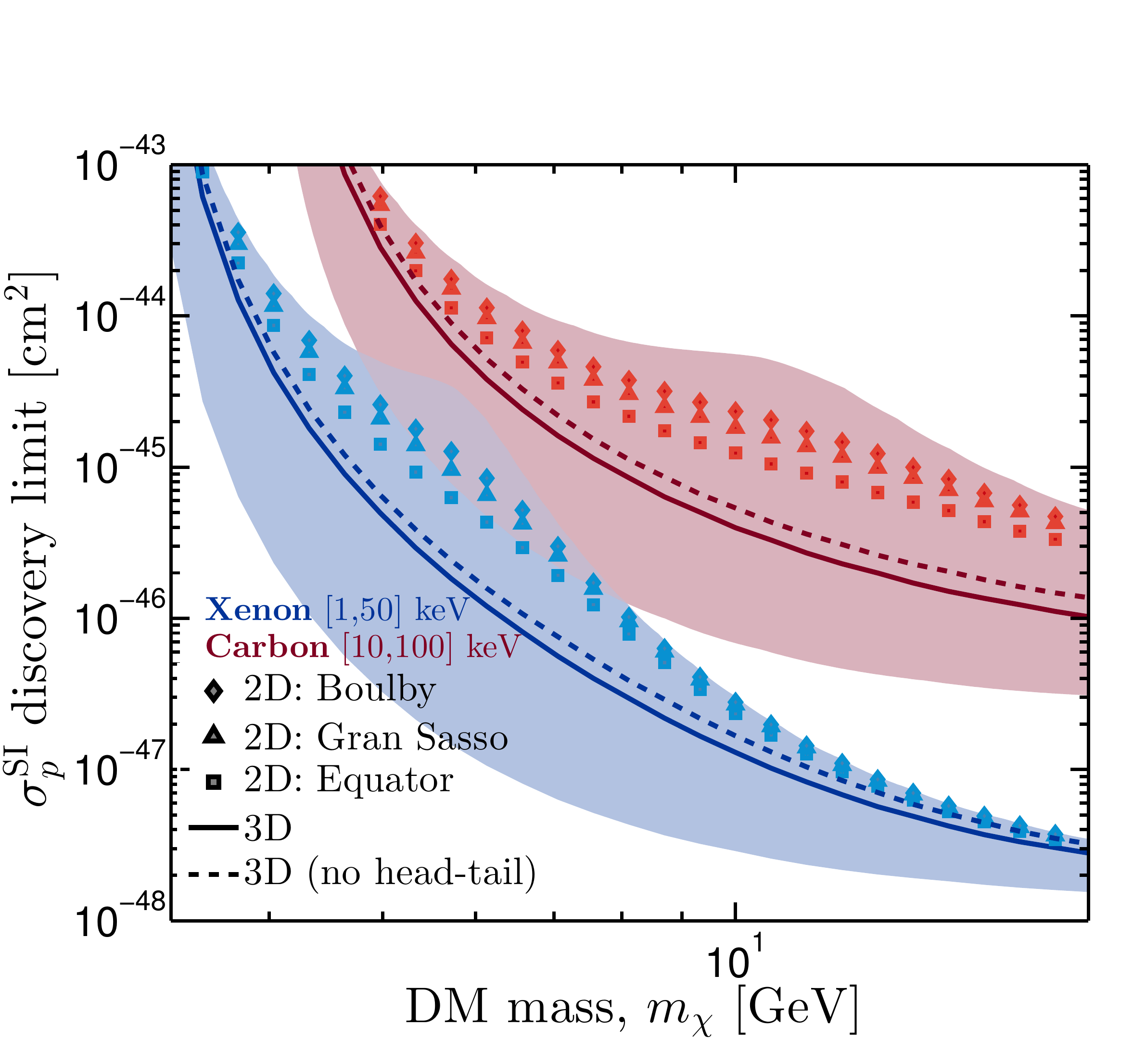}
\includegraphics[width=0.49\textwidth]{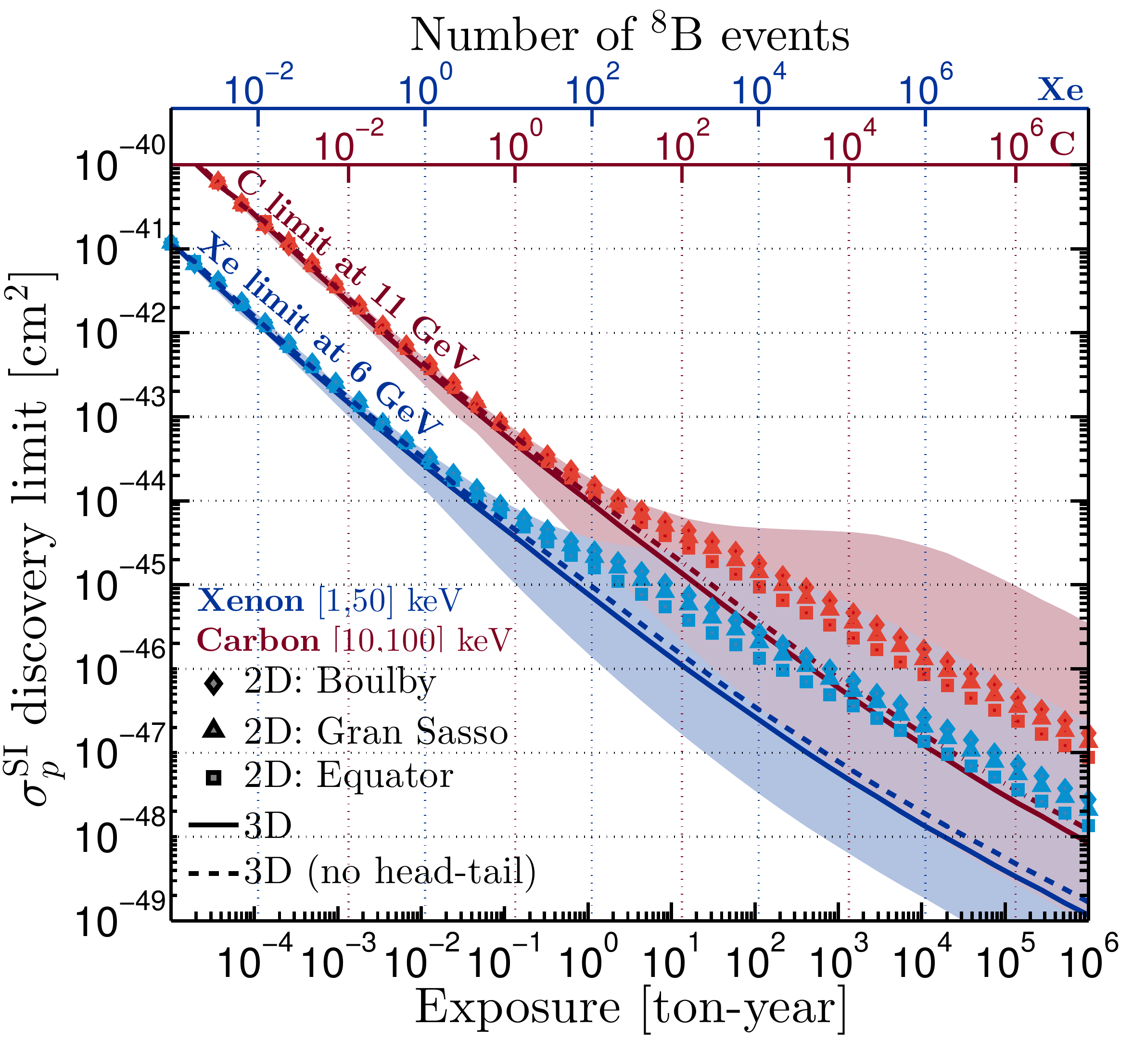}
\caption{SI discovery limits for a xenon (blue) and carbon (red) experiment in the presence of neutrino backgrounds. The markers show the discovery limits for 2D experiments located at Boubly ($\filledmedlozenge$), Gran Sasso ($\blacktriangle$) and the equator ($\filledmedsquare$). The solid and dashed lines without markers are for full 3D experiments but in the latter case with no head-tail sensitivity. The blue and red shaded regions show the difference between the discovery reach of an idealised non-directional experiment (upper edge) and an idealised directional experiment with time resolved 3D readout and head-tail measurement (lower edge). In the left-hand panel we show the discovery limits as a function of DM mass for fixed detector exposure: 10 ton-years for xenon and 100 ton-years for carbon. In the right-hand panel we display the evolution of the fixed DM mass discovery limit with increasing exposure: 6 GeV for xenon and 11 GeV for carbon.}
\label{fig:NuFloor_2DHT}
\end{figure*}
In Fig.~\ref{fig:NuFloor_2DHT} we show the DM mass (left) and detector exposure (right) dependence of the SI discovery limits in the presence of neutrino backgrounds (as in Figs.~\ref{fig:NuFloor_WIMPMass}~and~\ref{fig:NuFloor_DetectorMass}). We again show discovery limits for 2D detectors at three locations as well as the discovery limit for a 3D detector with no head-tail sensitivity. In most cases we obtain qualitatively similar results to the isotropy and median recoil direction tests in Secs.~\ref{sec:nonideal-iso} and~\ref{sec:nonideal-med}. The 2D discovery limits suffer a loss in sensitivity of just over an order of magnitude compared with a 3D time-integrated experiment, and around two orders magnitude compared with an idealised directional experiment with Cygnus tracking. The 2D limits also evolve towards a $1/\sqrt{N}$ Poisson background regime with around 100 $^8$B neutrino events, much sooner than for 3D recoils. Upon removing head-tail sensitivity we see dramatically different overall effects to the previous tests. For both xenon and carbon the removal of sense recognition has a very small effect on the overall evolution of the limit. This is because the $^8$B and DM recoil distributions (which are separated by between 60$^\circ$ and 120$^\circ$, depending on the date) still exhibit a marked separation even after the forward-backward folding of the recoil sky.

We conclude here by emphasising that while head-tail sensitivity is crucial for distinguishing a DM signal from an isotropic background using model independent tests, it is much less important for distinguishing DM recoils from a background which has a very different anisotropic angular distribution, such as Solar neutrinos. We have also shown that for existing nuclear emulsion detectors which are only sensitive to projections of recoil tracks in the plane of the plates, reasonable discrimination between isotropic backgrounds and Solar neutrinos can be made but with a non-negligible loss in discovery capabilities compared with a full 3D experiment.

\section{Conclusions}
\label{sec:conclusions}
\begin{table*}[t]\centering
\ra{1.3}
\begin{tabularx}{\textwidth}{l|YY|YY|YY|YY}
\hline\hline
\multirow{2}{*}{\bf Test}   & \multicolumn{2}{c|}{\bf Cygnus-tracking} & \multicolumn{2}{c|}{\bf Time-int.} & \multicolumn{2}{c|}{\bf 2D (Boulby)} & \multicolumn{2}{c}{\bf No head-tail} \\
	    & {\bf Xe} & {\bf C } & {\bf Xe} & {\bf C} & {\bf Xe} & {\bf C} & {\bf Xe} & {\bf C}\\
\hline
Rejecting isotropy in 95\% of experiments at $3 \sigma$ & 39& 7& 76& 15 & 263 & 70 & $>10^5$ & 549\\
Rejecting isotropy in 95\% of experiments at $5 \sigma$ & 108& 18& 161& 38& 651 &160 &$>10^5$ & 1405\\
\hline
Median direction in 95\% of experiments to within 20$^\circ$ & 73& 10 & 115& 30& 427 & 110 & - & -\\
Confirm median direction in 95\% of experiments at $3 \sigma$ & 1706& 220& 2332& 689& $>10^5$ & $>10^5$ & - & -\\
\hline
SI discovery limits an order of magnitude below $\nu$-floor & 47 & 44& 350 & 426 & 9836 & 11644 & 526 & 638 \\
\hline\hline
\end{tabularx}
\caption{Summary of the number of events required for each of the results presented in Secs.~\ref{sec:ideal} and~\ref{sec:nonideal}. The first four rows display the number of required signal events for a 50 GeV DM particle. For the final row we display the number of observed $^8$B neutrino events required to reach a cross section an order of magnitude below the `neutrino floor' (the plateau of Fig.~\ref{fig:NuFloor_DetectorMass}) at masses of 6 GeV for xenon and 11 GeV for carbon. In each column we show the number of required events for our benchmark xenon and carbon experiments considering (from left to right): an ideal 3D Cygnus-tracking experiment, an ideal 3D time-integrated detector, a 2D time-integrated detector located at Boulby, and a 3D detector with no head-tail sense recognition. For the isotropy and median direction tests we set a 1 (40) keV threshold on the xenon (carbon) experiment, which requires a 2.9 kg-year (5.9 ton-year) exposure to observe 1 signal event from DM with a cross section of $\sigma^{\rm SI}_p = 10^{-45}$~cm$^2$. For the neutrino floor test we lower the carbon threshold to 10 keV to observe the neutrino background. In this case 1 $^8$B event in xenon (carbon) requires an exposure of 0.011 (0.13) ton-year. }
\label{tab:results}
\end{table*}

We have studied the discovery reach of time-integrated directional DM detectors. We summarise our quantitative results in Table~\ref{tab:results}. These results indicate that a directional detector without timing information should still possess some of the unique benefits of directional detection. For the initial goal of rejecting an isotropic background, we find that the sensitivity loss only amounts to an increase of around a factor of 2 in the number of required signal events (at both 3 and 5$\sigma$). We find similar results for the significance associated with measuring the median recoil direction. The expense of mounting a detector (and its shielding) on an equatorial telescope so that it can track Cygnus is therefore likely not warranted. Only in experiments that can reorient recoil events in the Galactic rest frame can the Galactic origin of such events be truly established. Nevertheless, given that there appear to be no known backgrounds that mimic the Southward preference of the time-integrated angular signature~\cite{Mei:2005gm}, a confirmation of the expected median direction by a time-integrated experiment would still provide very strong evidence for dark matter. 

We have performed the analysis for two benchmark target nuclei with different energy thresholds. The recoil spectrum is more anisotropic and focused towards $-\mathbf{v}_\mathrm{lab}$ for higher recoil energies. Therefore because we assume a lower energy threshold for the xenon experiment a larger number of signal events are required to achieve the same significance as in the higher threshold carbon experiment. A larger exposure is required to accumulate a given number of events in a high threshold experiment however. Model dependent likelihood techniques that incorporate recoil energy as well as direction~\cite{Billard:2009mf} would take into account the changing degree of anisotropy as a function of energy.  However the frequentist hypothesis tests have an advantage in that they do not require any parameterisation or fitting, simply relying on basic assumptions about the properties of the underlying angular recoil spectrum. 

In the case of neutrino backgrounds, we find that time-integrating detectors are still a very powerful approach for circumventing the `floor' faced by non-directional experiments. To reach cross sections an order of magnitude below the floor an exposure a factor of $7 - 10$ larger is required than for an experiment that measures the times and directions of the recoils. So although we note that time integration does significantly affect the evolution of the discovery limits with event number, as shown in Fig.~\ref{fig:NuFloor_DetectorMass}, crucially these limits do not suffer from the same background saturation exhibited by the non-directional limits. Once one considers the very large detector exposures required to even observe the neutrino background, it could be argued that nuclear emulsion detectors are in fact the most promising strategy to deal with the neutrino background, since scaling gaseous detectors to similar target masses requires prohibitively large and expensive volumes. Although we study only the low mass Solar neutrino floor - since it is of immediate relevance to the upcoming generation of experiments - we expect that an analysis of the neutrino floor due to atmospheric neutrinos and the DSNB would find qualitatively similar results as in Ref.~\cite{O'Hare:2015mda}. We omit this analysis because exposures in excess of 1000 ton-years are required to observe these neutrino backgrounds, and the benefit afforded by directionality in discriminating them from heavier DM masses is not as impressive (even for idealised detectors).

The above conclusions are for detectors with 3D readout that can measure the senses of recoil tracks. We also considered 2D readout and detectors without head-tail sensitivity (as is presently the case for NEWSdm).  For rejecting isotropy and confirming the median direction we show that only measuring 2D recoils is not a major disadvantage (cf. Ref.~\cite{Billard:2014ewa}). In particular, if it turns out that the experiment can be tilted to mimic an equator-based experiment then the lack of the vertical track component would prove to be essentially unimportant. This would require the nuclear emulsion detector to be stable over a timescale of several years while tilted at a fixed angle. If this is the case then a fixed tilted set up would be advantageous as it would  be cheaper and more feasible than a rotating equatorial mount. Furthermore, the sensitivity is not significantly less than for an experiment which can measure the recoil tracks in 3D. Instead of trying to achieve 3D readout, it would be far more beneficial to search for head-tail effects which would allow the sense of the recoils to be measured. In agreement with previous work, e.g. Ref.~\cite{Morgan:2004ys}, we reiterate that head-tail sensitivity is of crucial importance for detecting dark matter. Without it the number of events required to reject isotropy is increased by orders of magnitude and measuring the median direction is practically impossible. On the other hand, for subtracting the neutrino background in non-ideal detectors we come to a slightly different conclusion. There we find head-tail sensitivity is much less important due to the stark differences in the angular distribution of the Solar neutrino and DM recoil spectra. We also find that a 2D detector can still make excellent progress past the neutrino floor, particularly at DM masses that are most saturated by the background (c.f. Ref.~\cite{O'Hare:2015mda}).

We have ignored the effects of astrophysical uncertainties in this work. However, it has been shown in the past that directional detectors are best suited for measuring the velocity distribution and probing local DM astrophysics~\cite{Lee:2012pf,Kavanagh:2016xfi,Laha:2016iom,Nagao:2017yil}. Analogous tests to those described here have been used to evaluate the detectability of anisotropies and substructure in the velocity distribution, hence one could combine our analysis with those of previous works, e.g.~Refs.~\cite{Morgan:2004ys,O'Hare:2014oxa}. Unfortunately we anticipate that the effects of time integration may be too severe. A crucial requirement for performing `DM astronomy' is that recoils can be oriented in the Galactic frame. Very large degrees of anisotropy may be detectable but it is likely that many thousands of events would be required to make concrete statements. While fully time-resolved 3D directional experiments may be essential for probing the fine structure of the DM velocity distribution, we have shown here that time-integrated directional detectors should still be powerful tools for confirming the DM origin of a signal and for eliminating the neutrino floor.

\acknowledgments{CAJO is supported by a Leverhulme Trust Research Leadership Award. BJK is supported by the European Research Council ({\sc Erc}) under the EU Seventh Framework Programme (FP7/2007-2013)/{\sc Erc} Starting Grant (agreement n.\ 278234 --- `{\sc NewDark}' project). AMG  acknowledges  support  from  STFC  grant ST/L000393/1.}

\appendix

\section{Neutrino backgrounds}
\label{nu}
\begin{table*}\centering
\begin{tabularx}{\textwidth}{Y|YYY|Y}
\hline\hline
$\mathbf{\nu}$ \bf{type}  & $E_{\nu}^{\rm{max}}$ (MeV) & $E_{r_{\rm{Xe}}}^{\rm{max}}$ (keV) & $E_{r_{\rm{C}}}^{\rm{max}}$ (keV) & $\Phi\pm \delta\Phi$ $\mathbf{(\rm{cm^{-2} \, s^{-1}})}$\\
\hline
${}^{8}\rm{B}$ & 16.36 & 4.494 & 47.41 & $\left(5.16\pm 0.11\right) \times 10^6$\\
$hep$ & 18.78 & 5.782 & 62.47 & $\left( 8.04\pm 2.41 \right) \times 10^3$\\
 DSNB & 91.20 & 136.1 & $3.212\times 10^3$ &$ 85.7\pm 42.7$\\
Atm. & 981.7 & $15.55\times 10^3$ & $1.45 \times 10^5$ & {$10.54\pm 2.1$}\\
\hline \hline
\end{tabularx}
\caption[Total neutrino fluxes with corresponding uncertainties]{The maximum neutrino energy, maximum recoil energies for xenon and carbon targets, and the neutrino flux and its uncertainty for the dominant neutrino backgrounds: $^8$B, $hep$, diffuse supernovae background (DSNB) and atmospheric (Atm.) neutrinos.
\label{tab:neutrino}}
\end{table*}

We only consider the neutrino background from coherent neutrino-nucleus elastic scattering (CNS). CNS proceeds via a neutral current, and as shown by Freedman~\cite{Freedman:1973yd} and subsequently Drukier \& Stodolsky~\cite{Drukier:1983gj} has a coherence effect at low momentum transfer that approximately scales with the number of neutrons squared. It was recently observed for the first time by the COHERENT experiment~\cite{Akimov:2017ade}. At higher recoil energies, generally above a few tens of keV, the loss of coherence is described by the nuclear form factor $F(E_r)$, for which we again use the standard Helm ansatz which is an excellent approximation at these low energies~\cite{Vietze:2014vsa}. The differential cross section as a function of the nuclear recoil energy ($E_r$) and neutrino energy ($E_\nu$) is given by 
\begin{equation}
  \frac{\textrm{d} \sigma}{\textrm{d}E_r}(E_r,E_\nu) = \frac{G_F^2}{4 \pi} Q^2_W m_N \left(1-\frac{m_N E_r}{2 E_\nu^2} \right) F^2(E_r) \,,
\end{equation}
where $Q_W = A-Z - (1-4\sin^2\theta_W) Z$ is the weak nuclear hypercharge of the nucleus, $G_F$ is the Fermi coupling constant, $\theta_W$ is the weak mixing angle and $m_N$ is the target nucleus mass. We assume CNS to be a pure standard model interaction and do not consider any exotic mediators as in, for example, Refs.~\cite{Cerdeno:2016sfi,Bertuzzo:2017tuf}.

The directional cross section can be written by noting that the scattering has azimuthal symmetry about the incoming neutrino direction so $\rm{d}\Omega_\nu = 2\pi \,\rm{d} (\cos\beta)$ and imposing the kinematical expression for the scattering angle, $\beta$, between the neutrino direction, $\hat{{\bf q}}_\nu$, and the recoil direction, $\hat{{\bf q}}_r$,
\begin{equation}
 \cos\beta = \hat{{\bf q}}_r \cdot \hat{{\bf q}}_\nu = \frac{E_\nu + m_N}{E_\nu}\sqrt{\frac{E_r}{2 m_N}} \,,
\end{equation}
with $\beta$ in the range $(0,\pi/2)$, using a delta function,
\begin{equation}
  \frac{\textrm{d}^2 \sigma}{\textrm{d}E_r \textrm{d}\Omega_r} = \frac{\textrm{d} \sigma}{\textrm{d}E_r} \, \frac{1}{2 \pi}\, \delta\left(\cos\beta - \frac{E_\nu + m_N}{E_\nu} \sqrt{\frac{E_r}{2 m_N}}\right) \,.
\end{equation}
The CNS event rate per unit detector mass, as a function of the recoil energy, direction and time, is given by the convolution of the double differential CNS cross section and the directional neutrino flux,
\begin{equation}
\frac{\textrm{d}^2 R(t)}{\textrm{d}E_r \textrm{d}\Omega_r} = \frac{1}{m_N} \int_{E_{\nu}^\textrm{min}} \frac{\textrm{d}^2 \sigma}{\textrm{d}E_r\textrm{d}\Omega_r} \frac{\textrm{d} \Phi(t)}{\textrm{d}E_\nu \textrm{d}\Omega_\nu } \textrm{d}E_\nu \,. 
\label{eq:isonu}
\end{equation}

Due to the eccentricity of the Earth's orbit, the Earth-Sun distance has an annual variation leading to a modulation in the Solar neutrino flux as seen by an Earth-based experiment (e.g. Ref.~\cite{Agostini:2017iiq}) such that,
\begin{equation}
  \frac{\textrm{d}^2 \Phi(t)}{\textrm{d}E_\nu \textrm{d}\Omega_\nu}  =  \frac{\textrm{d} \Phi}{\textrm{d} E_\nu} \,\left[ 1 + 2 e \cos\left(\frac{2\pi(t- t_\nu)}{T_\nu}\right) \right] 
 \delta\left(\hat{{\bf q}}_\nu-\hat{{\bf q}}_\odot(t)\right) \,,
\label{eq:solarneutrinoflux_directional}
\end{equation}
where $t$ is the time from January 1st, $e = 0.016722$ is the eccentricity of the Earth's orbit, $t_{\nu} = 3$ days is the time at which the Earth-Sun distance is minimum (and hence the Solar neutrino flux is largest), $T_{\nu} = 1$ year, $\hat{{\bf q}}_{\nu}$ is a unit vector in the direction of interest and $\hat{{\bf q}}_\odot(t)$ is a unit vector in the inverse of the direction towards the Sun. We ignore the tiny angular spread in incoming neutrino directions due to the angular size of the Sun's core on the sky, see e.g. Ref.~\cite{Davis:2016hil}. On the other hand, the DSNB is expected to be isotropic and constant in time. The atmospheric neutrinos we model as isotropic since the weak enhancement towards the horizon~\cite{Battistoni:2005pd,Battistoni:2002ew} is almost entirely washed out after the stochastic scattering process~\cite{O'Hare:2015mda}. The overall normalisations for each neutrino flux, along with the maximum neutrino and neutrino-induced recoil energies, are given in Table~\ref{tab:neutrino}.

\bibliography{TimeAveraged.bib}

\end{document}